\documentclass[12pt,preprint]{aastex}
\newcommand{\av}{$A_V$}

\newcommand{\eleven}{$11.6$}
\newcommand{\elevenn}{$([11.6]-N)$}

\newcommand{\ergs}{erg\,s$^{-1}$}
\newcommand{\jhks}{$JHK_s$}
\newcommand{\jk}{$(J-K_s)$}
\newcommand{\keleven}{$(K_s-[11.6])$}
\newcommand{\kcit}{$K_{CIT}$}
\newcommand{\kms}{km\,s$^{-1}$}
\newcommand{\kn}{$(K_s-N)$}
\newcommand{\knphot}{$(K_s-N)_{phot}$}
\newcommand{\ks}{$K_s$}
\newcommand{\kstwelve}{$(K_s-[12])$}
\newcommand{\ktwelve}{$(K_{CIT}-[12])$}

\newcommand{\mirac}{MIRAC-BLINC}
\newcommand{\qs}{$Q_s$}
\newcommand{\vj}{$(V-J)$}

\slugcomment{To appear in the Astrophysical Journal}

\shorttitle{Tuc-Hor Debris Disk Survey}
\shortauthors{Mamajek et al.}

\begin{document}
\title{Constraining the Lifetime of Circumstellar Disks in the
Terrestrial Planet Zone: A Mid-IR Survey of the 
30-Myr-old Tucana-Horologium Association}

\author{Eric E. Mamajek, Michael R. Meyer, Philip M. Hinz, 
William F. Hoffmann}
\affil{Steward Observatory, 
Department of Astronomy, The University of Arizona,\\ 
933 N. Cherry Ave., Tucson, AZ 85721}

\author{Martin Cohen}
\affil{Radio Astronomy Laboratory, 601 Campbell Hall, University of 
California at Berkeley,\\
Berkeley, CA 94720}

\author{Joseph L. Hora}
\affil{Harvard-Smithsonian Center for Astrophysics,\\
60 Garden St., MS-65, Cambridge, MA 02138}

\email{eem@as.arizona.edu}

\begin{abstract}
We have conducted an N-band survey of 
14 young stars in the $\sim$30 Myr-old Tucana-Horologium
Association to search for evidence of warm, circumstellar
dust disks. Using
the MIRAC-BLINC camera on the Magellan I (Baade) 6.5-m
telescope, we find that none of the stars have a statistically
significant N-band excess compared to the predicted 
stellar photospheric flux. 
Using three different sets of assumptions, this null result
rules out the existence of the following around these
post-T Tauri stars:
(a) optically-thick disks with inner hole radii of $\lesssim$0.1\,AU,
(b) optically-thin disks with masses of $>$10$^{-6}$\,M$_{\oplus}$ 
(in $\sim$1-\micron-sized grains) within $\lesssim$10\,AU of these stars,
(c) scaled-up analogs of the solar system zodiacal dust cloud
with $>$4000$\times$ the emitting area. 
Our survey was sensitive to dust disks in the terrestrial
planet zone with fractional luminosity of 
log(L$_{dust}$/L$_*$) $\sim$ 10$^{-2.9}$, yet none were found. 
Combined with results from previous surveys, these data
suggest that circumstellar dust disks become so optically-thin
as to be undetectable at N-band before age $\sim$20\,Myr.
We also present N-band photometry for several members of
other young associations and a subsample of
targets that will be observed with {\it Spitzer Space Telescope} by
the {\it Formation and Evolution of Planetary Systems} (FEPS) Legacy 
Science Program. Lastly, we present an absolute calibration of MIRAC-BLINC
for four filters ($L$, $N$, $11.6$, and $Q_s$) on 
the Cohen-Walker-Witteborn system.

\end{abstract}


\section{INTRODUCTION}

Circumstellar disks appear to be a nearly ubiquitous by-product
of the star-formation process. 
Most low-mass stars in the youngest star-formation 
regions (e.g. the $\sim$1-Myr-old \object[ONC]{Orion Nebula Cluster}) 
have spectroscopic or photometric evidence of a 
circumstellar disk \citep{Hillenbrand98}.
The masses of
circumstellar disks found around some T Tauri stars 
are similar to that of the minimum
mass solar nebula \citep{Beckwith90}. Their physical
sizes are similar to that of our own solar 
system \citep[10s-100s AU;][]{McCaughrean96}.
Considering their masses, dimensions, and appearance
at the very earliest stages of stellar evolution,
these disks are considered ``protoplanetary''. Radial
velocity surveys of
nearby solar-type stars indicate that at least $\sim$5\% 
have at least one Jupiter-mass planet orbiting within
a few AU \citep{Marcy00}, indicating that the formation
of gas giant planets is one likely outcome 
of circumstellar disk evolution.

The incidence of inner protoplanetary accretion disks 
diminishes with age, being very common
at ages $<$1 Myr, and very rare at $>$10\,Myr. 
The fraction of low-mass stars with disks 
inferred by IR excess (in the L-band; 3.5\,\micron) 
diminishes with age, with half losing 
their inner disks ($\lesssim$0.1\,AU) 
by age $\sim$3 Myr \citep[e.g.][]{Haisch01a}. 
Using population statistics of pre-MS stars
in the \objectname[NAME TAURUS MOLECULAR CLOUD]{Taurus 
molecular clouds}, multiple studies
have demonstrated that the transition time for 
disks inside of $\sim$1\,AU to go from 
optically-thick to optically-thin is $\sim$10$^5$\,yr 
\citep{Skrutskie90,Wolk96}. 
While some studies argue that accretion terminates
by age $\sim$6\,Myr \citep[][]{Haisch01a}, recent
studies suggest that it may continue at
lower accretion rates around some stars 
until at least age $\sim$10\,Myr 
\citep{Muzerolle00,Mamajek02,Lawson02,Lawson04}.
There are
preliminary indications that disks may persist longer for
stars in associations which lack massive 
OB stars and for the lowest mass stars 
\citep[e.g.][]{Lyo03,Haisch01b}. 
Although understanding the evolution of accretion
disks has improved, the evolution of dust in the
terrestrial planet zone is still largely unexplored.

Although the phenomenon of optically-thick accretion disks
appears to be isolated to the first $\sim$few Myr of a star's life,
numerous examples of older stars with optically-thin dust disks
have been found over the past two decades, primarily using
space-based IR telescopes, e.g. {\it InfraRed Astronomical
Satellite} (IRAS) and {\it Infrared Space Observatory} (ISO) 
\citep{Backman93,Lagrange00}. 
Optically-thin disks have been found around
stars over a wide range of ages and masses, but those with
the highest fractional luminosity ($f_d$\,=\,L$_{disk}$/L$_{\star}$)
are mostly confined to those younger than $<$few$\times$100 Myr 
in age \citep{Habing01,Spangler01}. 
These dusty ``debris'' disks are
inferred to be created by the collisions of larger bodies, rather
than primordial ISM dust \citep{Harper84,Backman93}. 
For micron-size dust grains orbiting between $\sim$0.1-10\,AU, 
the timescale for Poynting-Robertson drag to pull the grains 
into the star ($\sim$10$^{1-5}$ yr) is short compared to typical
stellar ages ($\sim$10$^{7-10}$ yr), implying that either the 
observed phenomena is short-lived, or that grains must be replenished
through collisions of larger bodies. There is preliminary evidence 
for a monotonic decrease in dust disk optical depth with age
\citep{Spangler01}, or possibly a more precipitous drop in optical
depth after age $\sim$400\,Myr \citep{Habing01}. 
Most of the known debris disks have been identified
by excess far-IR emission above that of the stellar 
photosphere \citep[e.g.][]{Silverstone00},
with characteristic dust temperatures of $\sim$30-100\,K. 
Despite efforts to find warm (T $\sim$ 200-300\,K) 
dust disks around field stars, precious few examples with 
detectable 10-12\micron\ excesses are known \citep[][]{Aumann91}. 

Observational constraints on the evolution of circumstellar
dust in the terrestrial planet zone are currently scarce. 
Planned observations with the recently launched {\it Spitzer 
Space Telescope} (SST)
by the {\it Formation and Evolution of Planetary Systems} 
(FEPS) Legacy Science program\footnote{http://feps.as.arizona.edu},
among others, will remedy this situation. FEPS plans to 
systematically trace the evolution of circumstellar gas and 
dust around sun-like stars
between the epoch of optically-thick accretion
disks (ages\,$\sim$\,few\,Myr) to the epoch of mature planetary
systems \citep[ages\,$\sim$\,few\,Gyr;][]{Meyer02,Meyer04}. 

Although SST promises to provide a leap in our understanding
of the circumstellar environs of stars, we can address a basic 
question about disk evolution using 
currently available ground-based facilities. 
{\it How much dust remains within a few AU of young stars during 
the epoch of terrestrial planet formation?} 
We address this question through a mid-IR survey of a 
sample of young, low-mass stars with ages of $\sim$30 Myr:
the \objectname[T-HA]{Tuc-Hor Association}. 

Dynamical simulations suggest that, given the surface mass
density of the minimum-mass solar nebula, runaway growth
can take place and form Moon-sized planetary ``embryos'' 
within $\sim$10$^{5}$\,yr \citep[][]{Wetherill93}. 
When the largest embryos reach radii of $\sim$1000\,km,
gravitational interactions increase the eccentricities
and collision velocities of smaller planetesimals, causing
more dust-producing collisions \citep{Kenyon04}.
Over the next $\sim$10$^{7}$-10$^{8}$\,yr, the growth of
the largest embryos is dominated
by giant impacts, which consolidate the embryos into a
small number of terrestrial planets \citep{Agnor99,Chambers01}.
During this epoch in our own solar system, 
the proto-Earth is hypothesized to have been
impacted by a Mars-sized planetesimal, which formed the 
Earth-Moon system \citep{Hartmann75,Stevenson87}.
Chronometry studies using radioactive parent-daughter systems 
(such as $^{182}$Hf-$^{182}$W) suggest that the 
Earth-Moon impact occurred 25-35\,Myr after the formation of 
the solar system \citep{Kleine02,Kleine03}.
We know that around at least one star (our Sun),
terrestrial planets were forming at age $\sim$30\,Myr. 

In \S\ref{obs} we describe the sample, our mid-IR 
observations, and
data reduction. \S\ref{results} presents the results of our
photometry. We present three simple circumstellar disk
models in \S\ref{disk_models}, and calculate upper limits
to the amount of dust orbiting within $\lesssim$10 AU of
the stars observed. In \S\ref{discussion} we discuss
our results in light of previous observational and
theoretical efforts in order to better understand
disk evolution around young stars. In the Appendix, we
present information related to the photometric calibration
of the MIRAC-BLINC system, as well as details regardings 
stars for which mid-IR excesses were detected. 

\section{OBSERVATIONS \label{obs}}

\subsection{The Sample \label{sample}}

Our mid-IR survey contains 14 low-mass stars that we argue
are probable members of the $\sim$30-Myr-old \objectname[T-HA]{Tuc-Hor
Association}. The observations of the \objectname[T-HA]{Tuc-Hor} stars
are presented in Table \ref{tuc_obs}. Observations
of some \objectname[T-HA]{Tuc-Hor} candidates that we reject as members, and 
other young stars (some of which are FEPS SST Legacy Science targets)
are included in Table \ref{others_obs}. Here we discuss
some technical aspects of the \objectname[T-HA]{Tuc-Hor} sample. 

The \objectname[Tucana Association]{Tucana} and Horologium associations are young
stellar moving groups that were identified nearly contemporaneously by 
\citet[ZW00]{Zuckerman00} and 
\citet[TDQ00]{Torres00}. 
\citet[ZSW01]{Zuckerman01a} present an 
updated membership list and photometry, and suggest that the 
similar ages, kinematics, and positions of the 
\objectname[Tucana Association]{Tucana} and Horologium associations 
allow us to consider them a single group (``\objectname[T-HA]{Tuc-Hor}''). 
We adopt $\sim$30~Myr as a reasonable age estimate
for \objectname[T-HA]{Tuc-Hor} based on recent values in the literature
(see Table \ref{tuc_age}). 

The published membership lists for \objectname[T-HA]{Tuc-Hor} appear to be
somewhat subjective and contain some stars that are unlikely
to be members. 
In order to include stars that are plausibly members of Tuc-Hor
as part of our study, we used kinematics as the primary 
membership criterion (in addition to the other criteria
used in previous studies). For the combined \objectname[T-HA]{Tuc-Hor}
membership lists of TDQ00, ZW00, ZSW01,
we calculate membership probabilities using the equations
of \citet{deBruijne99}, the heliocentric space motion of the 
\objectname[Tucana Association]{Tucana} nucleus from ZW00, and the 
best long-baseline proper 
motions available at present \citep[preferably Tycho-2 
or UCAC2;][]{Hog00,Zacharias03}. We reserve rigorous
discussion of membership and kinematics for a separate 
future study.
For this study, we retain only those stars whose membership
we could not reject based on proper motion data. We find that 
$\sim$30\% of the stars proposed as members
of \objectname[T-HA]{Tuc-Hor} have proper motions inconsistent 
with the motion of the assumed Tucana ``nucleus'' 
\citep[centered on the $\beta$ Tuc mini-cluster;  
using space motion vector given by][]{Zuckerman00}. 
The published membership lists
appear to contain several field stars, a handful of
which we observed with MIRAC (Table \ref{others_obs})
before appreciating
that they were probable non-members. 
Care should be used by investigators employing samples
of recently-discovered, diffuse stellar associations 
(e.g. \objectname[T-HA]{Tuc-Hor}) for the study of age-dependent 
stellar phenomena.

We include in our \objectname[T-HA]{Tuc-Hor} sample the active 
dwarf \objectname[HIP 490]{HD 105} (= \objectname[HD 105]{HIP 490}). 
\objectname[HIP 490]{HD 105} is in the same region of sky 
($\alpha$,\,$\delta$(ICRS)\,= 00$^h$05$^m$,\,--41$^{\circ}$45$\arcmin$)
as the other proposed \object[T-HA]{Tuc-Hor} stars,
and has a Hipparcos distance of $d$ = 40\,pc. 
For our calculations we adopt the long baseline
Tycho-2 proper motion \citep{Hog00} and the Tuc-Hor
space motion vector from \citet{Zuckerman00}. 
In subjecting \objectname[HIP 490]{HD 105} to the same kinematic tests as the
other \object[T-HA]{Tuc-Hor} candidates, we are unable to reject its
membership. The calculated cluster parallax (25.3\,mas)
and predicted radial velocity (0\,\kms)
agree very well with the observed
trigonometric parallax \citep[24.9\,$\pm$\,0.9\,mas; ][]{ESA97}
and measured RV \citep[+1.7\,$\pm$\,2.5\,\kms; ][]{Wichmann03}.
The equivalent width of the \ion{Li}{1} $\lambda$6707 line 
\citep[165 m\AA;][]{Cutispoto02} is similar to that
for early-G-type members of $\sim$50\,Myr-old 
\object{IC 2602} and \object{IC 2391} clusters \citep{Randich01}, 
and stronger than that found in $\sim$120-Myr-old 
\object[M 45]{Pleiades} stars \citep{Soderblom93}. 
Finally, the X-ray luminosity of \object[HIP 490]{HD 105} 
\citep[log(L$_X$) = 29.4\,\ergs;][]{Cutispoto02} 
is similiar to what is expected for early-G
stars with ages of 10-100 Myr \citep{Briceno97}. Therefore, we 
conclude that \objectname[HIP 490]{HD 105} is a likely member of
the \object[T-HA]{Tuc-Hor} association.

\subsection{Data Acquisition \label{data_acq}}

Mid-IR images of the \object[T-HA]{Tuc-Hor} members and other young stars were 
obtained during the nights of 6-10 August 2001 and 22 August 
2002 (UT) with the \mirac\ instrument on the Magellan I 
(Baade) 6.5-m alt-az telescope at Las Campanas Observatory, Chile. 
The {\it Mid-InfraRed Array Camera} (MIRAC-3) contains a Rockwell 
HF16 128 $\times$ 128 hybrid BIB 
Si:As array, and was built at the Steward Observatory, University of
Arizona, and the Harvard-Smithsonian Center for Astrophysics 
\citep{Hoffmann98}. The {\it BracewelL Infrared Nulling 
Cryostat} (BLINC) is a nulling interferometer mated to 
MIRAC-3 \citep{Hinz00}. In our observing mode, however,  
BLINC is used as a re-imaging system, reducing the f/11 beam
from the Magellan tertiary mirror to a f/20 beam 
required for the MIRAC-3 instrument. The pixel scale
is 0.123\arcsec/pixel, resulting in a FOV of 15.7\arcsec.

Our intent was to survey for circumstellar dust surrounding
our target stars in the
terrestrial planet zone ($\sim$0.3-3\,AU), 
with characteristic temperatures 
of $\sim$300\,K, and corresponding Wien emission peak at
$\sim$10\,\micron. 
Observations were obtained with either the wide-band $N$ filter
\citep[$\lambda_{iso}$ = 10.34\micron\ for an A0 star, where
$\lambda_{iso}$ is the isophotal wavelength, e.g.][]{Golay74}
or narrow-band ``\eleven'' filter
($\lambda_{iso}$ = 11.57\micron\ for an A0 star)
in standard chop-nod mode 
\cite[4 position beam switching; see Appendix 1 of][]{Hoffmann99}. 
The nod and chop separations were 8\arcsec, 
and the chop (frequency of 3-10 Hz) was in a direction
perpendicular to the nod vector. 
The chop-nod imaging technique produces two 
positive and two negative images of the star in a 
square configuration on the detector.
The nod separation was chosen so
that all four images of the star appear on the
detector with sufficient room for determination of
the background flux in annuli surrounding each star image. 
We found that using chop frequencies between 3 and 10 Hz 
mitigated the effects of poorly subtracted
sky background (background noise increases as chop
frequency decreases) while maintaining observing efficiency 
(increasing the chop frequency adds overhead time,
with minimal improvement in background subtraction).
Chopping was done with an internal pupil plane 
beam-switching mirror within BLINC.
The frame time (on-chip integration time) was either 10\,ms ($N$-band) 
or 40\,ms (\eleven-band),
and these frames were co-added in 15-30\,s long integrations per
nod beam. We found
that derotating the \mirac\ instrument 
(i.e. freezing the cardinal sky directions on the detector)
in the Nasmyth port
during observations resulted in poorer background subtraction
compared to turning the derotation off. Hence, for the majority
of observations taken during these nights, the
instrument derotation was turned off. The ability to 
guide the telescope while derotating the instrument, was
not available
during our observing runs. Hence the telescope was not guiding
during most of our observations. This had negligible impact on the
achieved image quality, but limited our ability to reliably 
co-add data for faint sources. 

\subsection{Reduction \label{reduction}}

The MIRAC images were reduced using the custom program
{\it mrc2fts} \citep{Hora91} and IRAF\footnote{
IRAF is distributed by the National Optical Astronomy 
Observatories, which are operated by the Association of 
Universities for Research in Astronomy, Inc., under 
cooperative agreement with the National Science Foundation.
http://iraf.noao.edu}
routines. Flat fields were constructed from images
of high (dome) and low (sky) emissivity surfaces.
A median sky frame was produced and subtracted from the
individual (N $\simeq$ 10) dome frames. 
The results were then median combined to 
produce the final flat field. The pixel-to-pixel 
variation in sensitivity ($\sim$2\% r.m.s.) of the 
MIRAC detector is small enough that flat-fielding 
had negligible effect on our derived photometry.

Aperture photometry was derived using the IRAF {\it phot} package. 
We used aperture radii of either 0.62'' (5 px) or 1.23'' (10 px), 
depending on which flux had the higher S/N ratio after the
photometric errors were fully propagated (dominated
by sky noise for large aperture or uncertainty in aperture
correction for smaller apertures). The photometry
derived with aperture radii of 5 px and 10 px were consistent
within the errors for all of the stars observed.
The background level was determined by measuring the mean
sky value per pixel in an annulus centered on the star 
with inner radius 
1.85'' (15 pix) and outer radius 3.08'' (25 pix) for
subtraction.
The background annulus radii were chosen so as to sample
a negligible contribution of the star's PSF, but 
to avoid the PSFs from the other images of the same star.
For the faintest sources, an aperture radius of 5\,px was 
usually used, in which
case an aperture correction was applied to place
all photometry on the 10\,px system. The aperture corrections
were determined nightly using standard stars, and the
typical correction to the 5\,px aperture radius photometry
was --0.32\,$\pm$\,0.05\,mag. Photometric solutions 
(zero points and airmass corrections) were determined for 
every night of observations. The typical airmass corrections
at $N$-band were 0.1-0.2~mag\,airmass$^{-1}$. The conversion
between fluxes (in mJy) and magnitudes is simply
{\it mag$_{\lambda}$}(star) = 
--2.5\,log\{$f_{\lambda}$(star)/$f_{\lambda}$(0)\}, where
$f_{\lambda}$(star) is the star's flux at wavelength $\lambda$,
and $f_{\lambda}$(0) is the flux of a zero-magnitude star
(see Appendix \ref{apx_filters}). The sensitivity
was such that we could detect
a star with of magnitude 7.5 in $N$-band, at 
S/N\,$\simeq$\,5, with 600\,s of on-source integration time. 

We observed standard stars taken from the 
MIRAC manual \citep{Hoffmann99}, the list of ESO IR standards 
\citep{vanderBliek96}, and the list of \citet{Cohen99}.
M. C. calculated an independent calibration of
the MIRAC photometric system, using the approach identical
to that described in \citet{Cohen03b}. Discussion on the input
data for the absolute photometric calibration, the
zero-magnitude attributes of the MIRAC filter systems,
and standard star fluxes, are given in
Appendix \ref{apx_filters}. The
absolute accuracy of the standard star fluxes among the
four filters ranges from 1.7-4.5\%.

\section{RESULTS \label{results}}

\begin{figure}
\epsscale{1.0}
\plotone{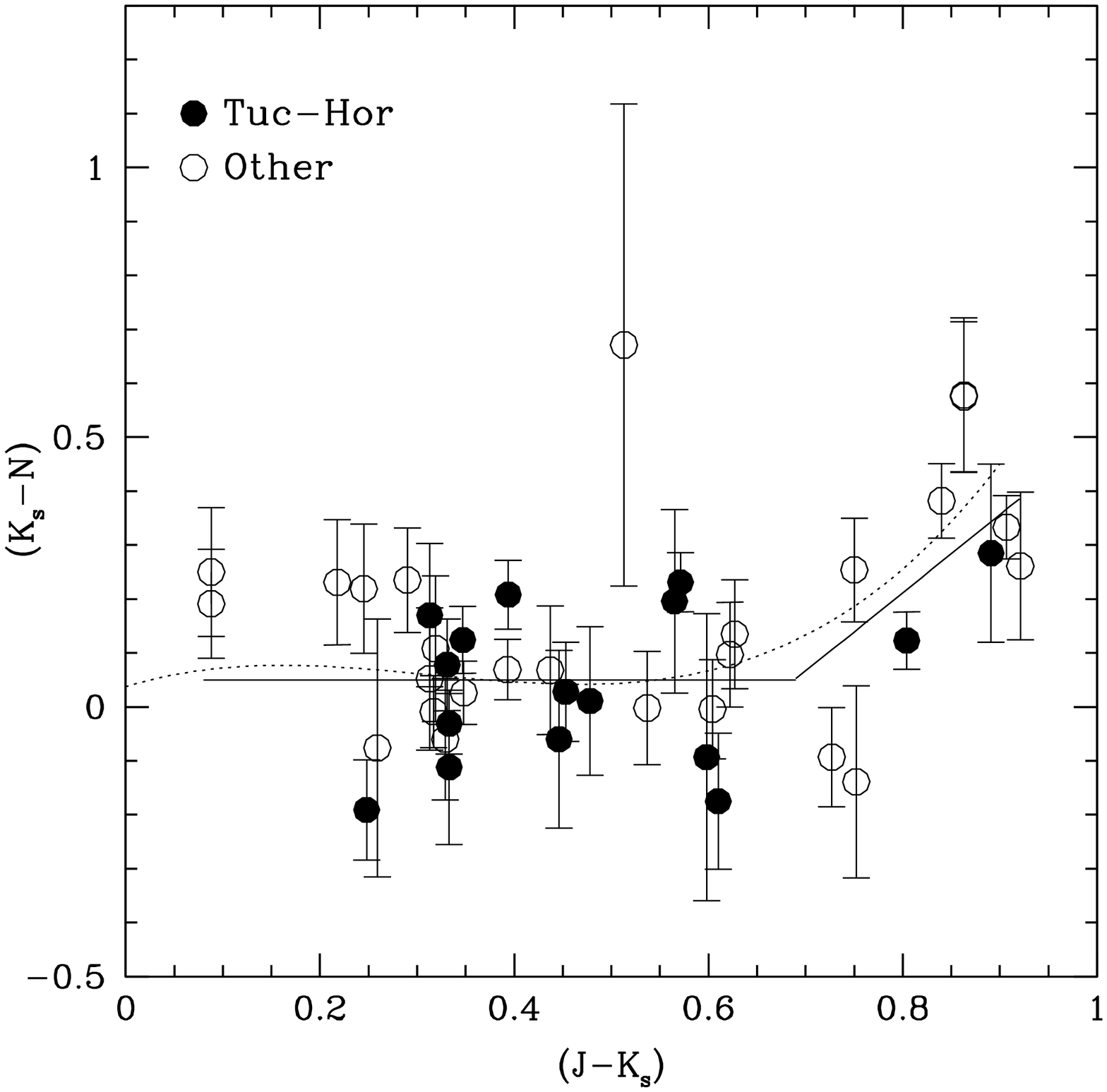}
\caption{\label{fig:colorcolor} Color-color diagram for 
Tuc-Hor members ({\it filled circles}) and other 
stars observed in this study ({\it open circles}). Two stars
(not members of Tuc-Hor) with 
significantly red \kn\ colors are not shown
(HD 143006 and J161411). 
The {\it solid line} is our 
adopted photosphere color relation (\S\ref{results}). 
The {\it dashed line} is a smoothed fit to the \ktwelve\ dwarf 
sequence of \citet[][]{Kenyon95}, where we calculate
\kn\ $\simeq$ \ktwelve$_{KH95}$ - 0.06.
}
\end{figure}

$N$-band photometry for young stars in \object[T-HA]{Tuc-Hor} 
is presented in Table \ref{tuc_phot}, while photometry for stars in other
regions (most belonging to young, nearby associations)
is presented in Table \ref{others_phot}. Near-infrared (\jhks)
photometry from
the 2MASS catalog \citep{Cutri03} was used to help predict the
brightness of the stellar photospheres at 10\micron. 
For the range of spectral types investigated, models
predict that \elevenn\,$\simeq$\,0.00 within our photometric 
errors (typically $\sim$0.05-0.10\,mag), 
hence we plot \kn\ and \keleven\ on the same
color-color plot, and generically
refer to these colors as ``\kn'' throughout. 
A color-color plot of the \object[T-HA]{Tuc-Hor} stars is 
illustrated in Fig. \ref{fig:colorcolor}. 

\notetoeditor{Figure \ref{fig:colorcolor} should go around here}

Fig.  \ref{fig:colorcolor} indicates that \kn\ colors are fairly uniform
for stars with \jk\ $<$ 0.7, i.e. for FGK stars.
We decided to analyze the M stars separately from the FGK stars.
We surmise that much of the structure in
the published color-color relations for dwarfs is probably 
due to statistical fluctuations
\citep[e.g. ][]{Cohen87,Waters87,Mathioudakis93,Kenyon95}. 
After examining our data and those from previous studies
of large dwarf samples, we decided to assume a 
constant \kn\ color for the photospheres of FGK stars. 
In calculating a mean intrinsic \kn\, color, we include
the FGK stars in Tables \ref{tuc_phot} and \ref{others_phot},
but exclude the FGK stars \object{HD 143006}, 
\objectname[PZ99 J161411.0-230536]{[PZ99] J161411} (both 
with \kn\,$\simeq$\,2), 
and \object{HIP 95270} and \object{HIP 99273} \citep[known to 
have far-IR excesses which may contaminate $N$-band flux, 
e.g.][]{Zuckerman04}.
The median, unweighted mean, and weighted mean \kn\ color 
for the FGK stars are all similar (0.05, 0.04\,$\pm$\,0.02,
and 0.07\,$\pm$\,0.02, respectively). 
These estimates agree well with the mean FGK dwarf color found by 
\citet[][ \kstwelve\,$\simeq$\,+0.04 
implies \kn\,$\simeq$+0.05]{Fajardo00} 
and are close to the value for AFGK-type dwarfs found by 
\citet[][ \ktwelve\,=\,+0.02 implying \kn\,=\,+0.01]{Aumann91}
\footnote{Although not explicitly stated, the K 
photometry from \citet{Aumann91} appears to be 
on the CIT system, where 2MASS~\ks\ = \kcit\ -- 0.019 
\citep{Carpenter01}. The MIRAC $N$-band photometric 
system assumes N\,=\,0.00 for \object{Vega}, 
whereas \citet{Aumann91} list [12] = +0.01 for \object{Vega}. 
We ignore any color terms in converting [12] to 
a predicted $N$ magnitude, and derive 
\kn\,$\simeq$\,(\kcit--[12]) -- 0.01.}.
Within the uncertainties, {\it our measured mean \kn\, color
for FGK dwarfs is consistent with previous determinations.
The mean colors for the young stars in our observing program
do not appear to be biased toward red \kn\ colors due to
the presence of circumstellar material.}
We adopt \knphot\,=\,+0.05 as the photospheric color
for FGK-type stars with \jk\,$<$\,0.69.

The observed \kn\ colors of the M-type stars are systematically 
redder than those of the FGK stars, as well as the M-giant
standards. 
We looked for independent confirmation that the turn-up in 
\kn\ color for the coolest dwarfs is a real effect,
and not due to circumstellar material. 
We measure a color of \keleven\,=\,0.33\,$\pm$\,0.06 for the
$\sim$12-Myr-old M0 star \object{GJ 803}. 
\citet{Song02} studied the spectral 
energy distribution of \object{GJ 803} and concluded that the 
observed cold dust excess detected by IRAS at 60\micron\ does not 
contribute significant flux at 12\micron, and that
the IRAS 12\micron\ flux is consistent with a NextGen
model photosphere. Our observed 11.6\micron\ flux 
(608\,$\pm$\,32 mJy) agrees very well with the 
predicted photospheric flux (633 mJy at 11.6\micron), 
as well as the color-corrected IRAS 12\micron\ flux 
(537\,$\pm$\,32 mJy). 
These observations confirm that the observed \keleven\
color for \object{GJ 803} is photospheric, and
that the turn-up in \kn\ color for M stars is real. It appears
that none of the M-type stars has a statistically 
significant mid-IR excess. 
For the purposes of calculating upper limits on mid-IR
excess, we fit a line to the mean \kn\ colors for the
KM-type stars with \jk\,$>$\,0.6 (excluding the 
T Tauri star \objectname[PZ99 J161411.0-230536]{[PZ99] J161411}), 
and model the M-type dwarf photosphere colors as: 
\knphot~=~\mbox{--0.947~+~1.448$\times$\jk}~~
\mbox{(0.69\,$<$\,\jk\,$<$\,0.92)}. 

We determine whether a star has detectable excess at $N$ or \eleven\
through calculating the excess as:

\begin{equation}
E(N)\,= N\,-\,N_{phot} = N\,-\,K_s\,+\,(K_s-N)_{phot}
\end{equation}
\begin{equation}
\sigma^2[E(N)]\,=\,\sigma^2[N] + \sigma^2[K_s] + \sigma^2[(K_s-N)_{phot}]
\end{equation}

\noindent The contribution to the $N$-band excess from interstellar
extinction will only become similar in size to our photometric 
errors if \av\ $\gtrsim$ 1-2\,mag, hence we can safely ignore
extinction for the stars observed. We examined the residuals 
(defined as $E(N)/\sigma [E(N)]$)
in order to identify statistically-significant outliers
(i.e. possible $N$-band excess stars). The $\sim$5-Myr-old
Upper Sco members \object{HD 143006} and 
\objectname[PZ99 J161411.0-230536]{[PZ99] J161411}
\citep{Preibisch99} both stand out with definite $N$-band excesses 
($\simeq$\,15-20$\sigma$), and
they are discussed further in Appendix B. 
There are two stars with positive 2-3$\sigma$ excesses (\object{HIP 6485} 
and \object{HIP 6856}), 
however there are three stars with 2-3$\sigma$ deficits (\object{HIP 9685}, 
\object{GJ 879}, and \objectname[PZ99 J161318.6-221248]{[PZ99] J161318}). 
Hence, the weak excesses for \object{HIP 6485} and \object{HIP 6856} are probably
statistical and not real. Excluding the two Upper Sco stars with
strong $N$-band excesses, we find that 56\,$\pm$\,12\% of the excesses
E(N) are within 1$\sigma$ of zero, and 88\,$\pm$\,15\% are within 
2$\sigma$ (uncertainties reflect Poisson errors). 
It does not appear necessary to introduce a non-zero uncertainty 
in the intrinsic \knphot. If one wanted to force 68\% of the
residuals to be within $\pm$1$\sigma$ and 95\% to be within
$\pm$2$\sigma$, then either $\sigma$[\knphot] $\simeq$ 0.07-0.09
mag, or our observational uncertainties are underestimated by $\sim$40\%.
We searched for, and could not find, a plausible reason why our
photometric errors would be underestimated by such a large amount.
The observations of our standard stars certainly do not support
a significant increase in our quoted photometric errors. 
More calibration observations
are required to see if this dispersion can be attributed to actual
structure in the intrinsic \kn\ colors of normal dwarf stars
as a function of spectral type.

Among the \object[T-HA]{Tuc-Hor} stars, only one star has as observed $N$-band flux 
$\geq$3$\sigma$ above that expected for stellar photosphere: 
\object{HIP 6856} (3.0$\sigma$ excess). However, there
is a Tuc-Hor member with a similarly sized flux {\it deficit} 
(\object{HIP 9685}; --2.9$\sigma$), so it is difficult to claim that the
excess for HIP 6856 is statistically significant.
{\it We find that none of the 14 \object[T-HA]{Tuc-Hor} members has an $N$-band 
excess more than 3$\sigma$ offset from the dwarf color relation}. 
We estimate a conservative upper limit to the $N$-band excess due to a 
hypothetical dust disk as 3$\times$ the uncertainty in the 
flux excess ($\sigma[E(N)]$; given in column 8 of Table \ref{tuc_phot}). 

\section{DISK MODELS \label{disk_models}}

Our survey was designed to be sensitive enough to detect the 
photospheres of young stars, hence we can place meaningful
constraints on the census of even optically-thin 
circumstellar disks in our target sample. We analyze the 
upper limits to possible mid-IR 
excess for the Tuc-Hor stars using three different models. 
The first model assumes a geometrically-thin, optically-thick
disk with a large inner hole. 
The second model assumes emission from an optically-thin disk
of single-sized grains. 
The third model treats the hypothetical disks as a scaled-up
version of the zodiacal dust disk in our solar system. 

\subsection{Optically-Thick Disk With Inner Hole \label{optthick}}

Infrared and submillimeter observations of T Tauri stars in dark 
clouds show that roughly half are orbited by an optically-thick 
circumstellar dust disk \citep[see review by][]{Beckwith99}. 
While the stars in our sample are roughly an order of magnitude 
older ($\sim$30 Myr) than typical T Tauri stars 
in dark clouds ($\leq$3 Myr), 
we can ask the question: If the \object[T-HA]{Tuc-Hor} stars have 
optically-thick, 
geometrically-thin disks, what is the minimum inner hole radius allowed 
by observations? \\

To answer this question for each star, we adopt the axisymmetric,
geometrically-thin, optically-thick disk model of \citet{Adams87}, 
and follow the formalism of \citet{Beckwith90}. 
While we assume a face-on orientation ($\theta$\,=\,0), our
results are not strongly dependent on this assumption. 
We also assume that 
the disk is optically-thick between $r_{in}$ and $r_{out}$ 
(300\,AU is assumed for all models) and at all frequencies 
(1 -- $e^{-\tau_\nu}$ $\sim$ 1). This is a safe assumption for 
T Tauri star disks in the wavelength regime probed in this study 
\citep[$\lambda$ $\leq$ 100\,\micron; ][]{Beckwith90}.\\
 
\begin{figure}
\epsscale{1.0}
\plotone{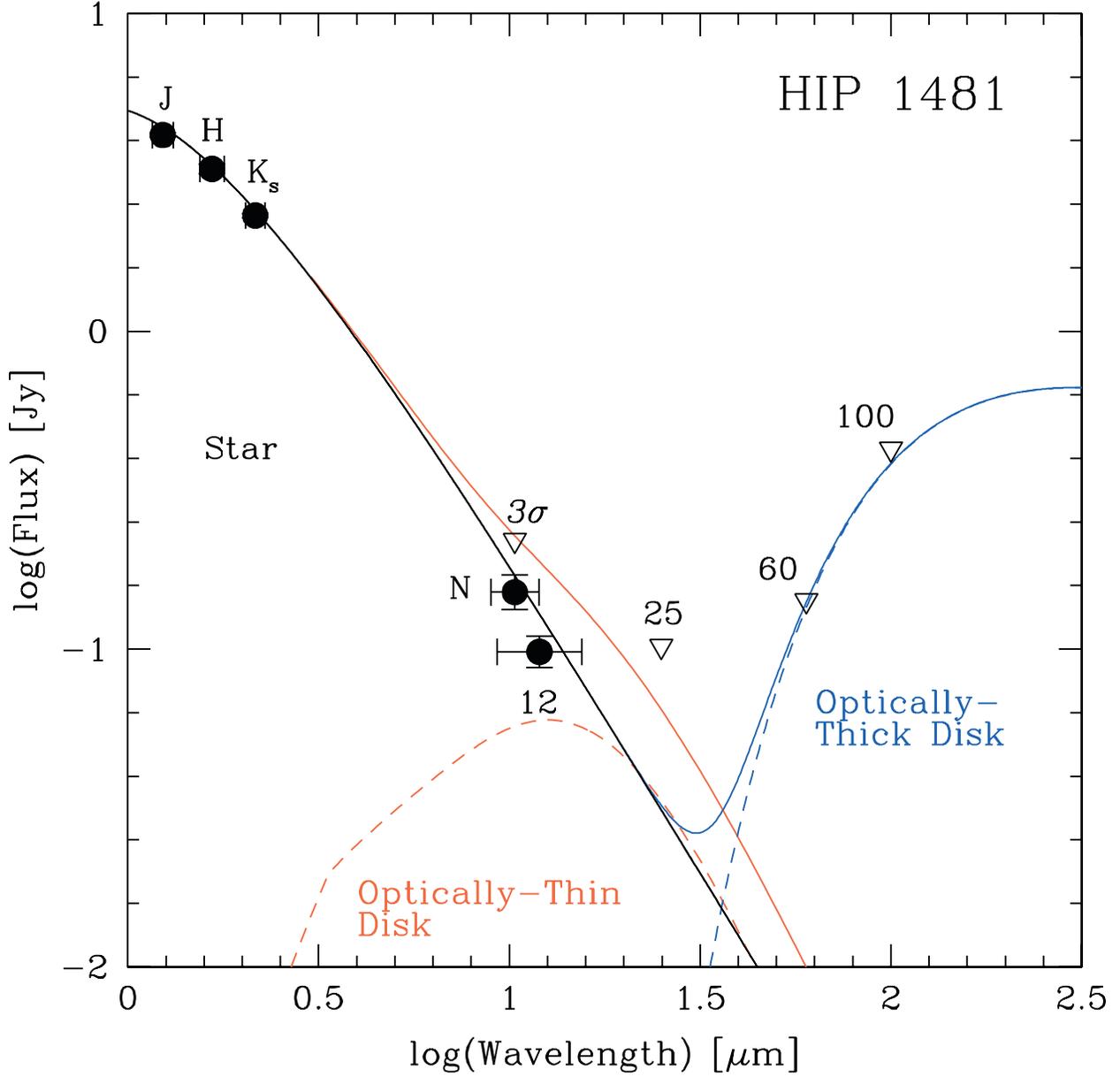}
\caption{\label{fig:sed}
Optically-thin and thick dust models fit to the MIRAC and IRAS photometry
for a typical Tuc-Hor star (HIP 1481). 
If the star has an optically-thick disk, its inner hole radius
must be $>$1.8\,AU (constrained by IRAS PSC 60\micron\ upper limits).
The stellar SED is approximated here as a 6026\,K blackbody.
The optically-thin disk model is conservatively matched to the 
3$\times$ the uncertainty in the $N$-band excess $E(N)$.
The kink in the spectral energy distribution for the optically-thin
model occurs at $\lambda$\,=\,2$\pi\bar{a}$ due to our
simple treatment of dust emissivity (\S\ref{optthin}).}
\end{figure}

Our $N$-band photometry alone allows us to rule out 
optically-thick disks with inner hole radii of $\sim$0.1 AU for the 
Tuc-Hor stars. Stronger constraints on inner disk radius for a hypothetical
optically-thick circumstellar disk can be calculated by including 
IRAS photometry.
For IRAS point sources, we adopt 25\,\micron, 60\,\micron, and 
100\,\micron\ fluxes and upper limits from the Faint Source 
Catalog \citep{Moshir90}. Where no IRAS point source is
detected, we adopt the IRAS Point Source Catalog upper limits 
of 0.5 Jy (25\,\micron), 0.6 Jy (60\,\micron), and 1.0 Jy 
(100\,\micron) \citep{IPAC86}. 
For the brightest stars, the IRAS 60\,\micron\ and 
100\,\micron\ data provide the strongest constraints 
on the existence of an optically thick disk, whereas
for the fainter K and M-type stars, the MIRAC photometry
provides the strongest constraint.  
IRAS did not map the region around \object{HD 105}, 
hence the inner hole radius we derive for this star is 
based only on the MIRAC $N$-band 3$\sigma$ upper
limit. 
In Fig. \ref{fig:sed}, we illustrate
a typical example (HIP 1481) of how the MIRAC and IRAS
photometry constrain the existence of optically-thick
disks around the \object[T-HA]{Tuc-Hor} stars.
The values we derive for the minimum inner hole radius
for a hypothetical optically-thick disk are given in column 5
of Table \ref{tuc_model}. The median value of the minimum 
inner hole radius is $\sim$0.3 AU (range: 0.1-7.9\,AU). 
The $N$-band and IRAS upper limits place the strongest constraints on 
inner hole size for the luminous F stars ($r_{in}$ $\gtrsim$ 5\,AU),
and the weakest constraints for the faint K/M stars 
($r_{in}$ $\gtrsim$ 0.1\,AU). 
{\it The MIRAC and IRAS photometry
easily rule out the existence of optically-thick disks
with inner hole radii of $\sim$0.1 AU of the $\sim$30 Myr-old 
\object[T-HA]{Tuc-Hor} stars.}

\notetoeditor{Figure \ref{fig:sed} should go around here}

\subsection{Optically-Thin Disk \label{optthin}}

In the absence of circumstellar gas, a putative mid-IR 
excess around a $\sim$30-Myr-old star 
would be most likely to be due to an optically-thin debris disk 
rather than an optically-thick T Tauri-type disk. 
The stellar ages 
($\sim$10$^{7.5}$\,yr) are orders of magnitude greater than 
the Poynting-Robertson drag timescale ($\sim$10$^{3-4}$\,yr) for 
typical interplanetary dust grains orbiting $\sim$1 AU from 
a solar-type star. Small dust grains must be continually 
replenished by collisions of larger bodies, or else they would 
be only detectable for astrophysically short timescales. 
Using a simple, single grain-size model, we
place upper limits on the amount of orbiting dust within
several AU of the $\sim$30-Myr-old \object[T-HA]{Tuc-Hor} stars.

Circumstellar dust grains surrounding young
main sequence stars should most likely have radii 
somewhere between the scale of typical ISM grains 
\citep[$\sim$0.01-1\,\micron; ][]{Mathis77}
and solar system zodiacal dust 
\citep[$\sim$10-100\,\micron; ][]{Grun85}. 
Theoretically, an ensemble of dust grains produced from a collisional
cascade of fragments is predicted to follow the equilibrium 
power-law size distribution \citep{Dohnanyi69}:
$n(a)\,$d$a$\,=\,$n_o$\,$a^{-p}$\,d$a$, where $p$\,=\,3.5.
Indeed this power law distribution is observed for
ISM grains \citep{Mathis77} and asteroids \citep{Greenberg89}.
With $p$\,=\,3.5, most of the mass is in the largest
(rarest) grains, but most of the surface area in the smallest 
(most common) grains. If the grain size 
distribution has a minimum cut-off, the mean grain size is 
calculated to be $\bar{a}$ = 5$a_{min}$/3 \citep[e.g.][]{Metchev04}. 
A limit on the minimum grain size $a_{min}$ 
can be estimated from consideration of radiation pressure blow-out 
\citep[e.g.][]{Artymowicz88}:

\begin{equation}
a_{min} = \frac{3 L_* Q_{pr}}{16 \pi G M_* c \rho} 
\end{equation}

\noindent Where $L_*$ is the luminosity of the star,
$Q_{pr}$ is the radiation pressure efficiency factor averaged over 
the stellar spectral energy distribution, $G$ is the
Newtonian gravitational constant, $M_*$ is the mass of the 
star, $c$ is the speed of light, and $\rho$ is the grain
density \citep[assumed to be 2.5\,g\,cm$^{-3}$;][]{Grun85}. 
The minimum grain size corresponds to the case where the ratio of 
the radiation pressure force to the stellar gravitational force is 
$F_{rad}$/$F_{grav}$ = 1. For this calculation
we assume the geometric optics case where $Q_{pr}$ = 1. For
the idealized grain orbiting the Sun at 1 AU, we calculate 
$a_{min}$\,=\,0.2\,\micron~and 
$\bar{a}$\,=\,0.4\,\micron. 
The grain size lower limit may be larger if
the momentum imparted by stellar winds dominates 
radiation pressure. The minimum grain size will be
somewhat lower if we calculate $Q_{pr}$ using Mie
theory and stellar spectral energy distributions,
instead of adopting the geometric optics case.
Highlighting the uncertainty in this calculation, we
note that the value of $\bar{a}$ that we calculate for the Sun 
is $\sim 10^2\times$ smaller than the mean interplanetary
dust particle orbiting in the Earth's vicinity \citep{Grun85}. 
This is largely due to a complex interplay between Poynting-Robertson 
drag and collisions. Increasing the cross-section
of dust particles in the solar system zodiacal dust
cloud by $\sim$10$^{4}$ (i.e. comparable to what we
are sensitive to in Tuc-Hor, see \S\ref{zody}), will decrease
the collision timescale, and correspondingly decrease the 
mean particle size to comparable to the blow-out grain size 
\citep{Dominik03}.

We model the thermal emission from an optically-thin 
disk of single-sized dust grains of radius $\bar{a}$ orbiting
in an annulus between inner radius $r_{in}$ and outer radius 
$r_{out}$. Spherical grains emit thermally at a temperature $T_{d}$ 
where the incident energy flux from the star is equal to the
isotropically emitted output energy flux of the grain. We approximate
the emissivity $\epsilon_\lambda$ of the single-size dust grains by 
using the simple model of \citet[][]{Backman93}:
emissivity $\epsilon_\lambda\,=\,1$ for 
$\lambda\,<\,2\pi a$, and 
$\epsilon_\lambda\,=\,(\lambda/2\pi a)^{-\beta}$ for longer 
wavelengths, where we assume $\beta$\,=\,1.5. Our adopted 
value of $\beta$ is similar
to that observed for zodiacal dust 
\citep[see Fig. 2 of][]{Fixsen02}
and ISM grains \citep{Backman93}.
The mass opacity is calculated as 
$\kappa_\lambda\,=\,3\epsilon_\lambda/4a\rho$.
The optical depth of emission through the disk annulus is 
$\tau_\lambda\,=\,\Sigma\kappa_\lambda$,
where $\Sigma$ is the surface density of the disk in g\,cm$^{-2}$. 
We calculate the orbital distance from the star ($r$) of dust grains heated
to temperature $T$ using 
equation \#5 from \citet{Wolf03}. We verify
that this relation is valid by comparing our 
calculations with \citet{Backman93} for grains much larger
than the Wien peak of incident light (blackbody case)
and for grains much smaller than the Wien peak of incident light
(e.g. ISM grains). For the \object[T-HA]{Tuc-Hor} stars, 
the Wien peak of incident starlight is comparable to 
$\bar{a}$.
We assume a flat mass surface density
profile ($\Sigma\,=\,\Sigma_o\,r_{AU}^{-p}$; $p\,=\,0$), 
which is appropriate for a population of dust grains 
in circular orbits subject to Poynting-Robertson drag 
\citep[see discussion in \S4.1 of][ and references therein]{Wolf03}. 
This predicted power law is close to what is observed for the 
zodiacal dust disk in our own solar system 
\citep[$p\,=\,0.34$;][]{Kelsall98}. 

Where exactly to define the inner and outer radii of a 
hypothetical dust disk requires some basic modeling. 
Among the \object[T-HA]{Tuc-Hor} stars, 90\% of the thermal emission 
from our hypothetical
disk model at $N$-band comes from within 
$\approx$1.5-2.2$\times r_{W}$ of the star, 
where $r_{W}$ is the radius at which the dust is at
the Wien temperature ($T_W$) for the isophotal wavelength of the 
$N$-band filter, and 
$T_{W}\,=\,2898\lambda^{-1}_{\micron} (5/(\beta+5))$ 
\citep[eqn. 6.9 of][]{Whittet03}. 
For simplicity,
we adopt a consistent definition of the outer radius for
all stars as 2$\times r_{W}$. Beyond 2$\times\,r_{W}$, 
the hypothetical dust disk contributes negligible flux 
($\lesssim$10\%) to what is observed in the $N$-band filter.  
Approximately 50\% of the thermal emission observed
in $N$-band from a hypothetical dust disk comes from within
$\lesssim$0.4-0.5$\times\,r_{W}$ of the star. 
For $r_{in}$, we adopt the radius for which the 
grain temperature is 1400\,K -- approximately the silicate 
dust sublimation limit. The temperatures of
the inner edges of typical T Tauri star disks
appear to be near this value \citep{Muzerolle03}. 
For the \object[T-HA]{Tuc-Hor} stars, $r_{in}\,\sim\,0.05$ AU and 
$r_{out}\,\sim\,10-15$ AU. Hence we are most sensitive
to dust at orbital radii comparable to our
inner solar system. 

Results for a typical set of fitted model parameters for our 
optically-thin disk model are illustrated in Fig. \ref{fig:sed}. 
Our calculations suggest that the
survey was sensitive to dust disk masses of 
$\sim 2\times 10^{-6}\,M_{\oplus}$
($\sim 10^{22}$\,g) in a single-sized dust grain population
(of uniform size $\bar{a}$, typically 0.1-2\,\micron). 
Our optically-thin model puts upper limits 
of $\Sigma_o$\,$\simeq$\,10$^{-6}$-10$^{-5}$ g\,cm$^{-2}$ on the surface
density of micron-sized dust grains in the $\sim$0.1-10\,AU
region around the \object[T-HA]{Tuc-Hor} stars. 
{\it For the masses and surface densities quoted, we assume that
all of the mass is in dust grains of size $\bar{a}$.}
In order to convey how sensitive our assumptions are for 
our final results, we show the effects of changing various 
parameters on our results in Table \ref{effects_of_change}. 
The dust disk masses that we calculate are similar
to those found by other studies \citep{Chen01,Metchev04} 
which also use the single grain-size approximation. The 
dust mass surface density upper limits that we calculate
are $\sim$10$^{-7}$-10$^{-6}\times$ that of the solids 
in the minimum mass solar nebula \citep{Weidenschilling77}, 
however we are not sensitive to bodies much larger than
the wavelength of our observations, or to gas. 

\subsection{Single-Temperature Zody Disk \label{zody}}

Another simple model to apply to our data is that of a 
scaled-up version of the terrestrial-zone zodiacal dust cloud in our 
own solar system. Though the detailed zodiacal dust model for the 
inner solar system is quite complex \citep{Kelsall98}, it can also be 
approximated by a single temperature blackbody (T\,=\,260\,K)
with bolometric luminosity 8\,$\times$\,10$^{-8}$\,L$_{\odot}$ 
\citep{Gaidos99}. 
This luminosity and temperature imply an equivalent surface area 
of 5\,$\times$\,10$^{-6}$\,AU$^2$ $\simeq$
1\,$\times$10$^{21}$\,cm$^2$. \citet{Gaidos99} defines this 
area as 1 ``zody'' (1 ${\cal Z}$). The unit is useful for 
comparing relative amounts of exozodiacal dust between the Sun 
and other stars.

With none of our stars having statistically significant $N$-band 
excesses, we calculate upper limits to the number of zodys present 
using 3$\times$ the uncertainty in the excess measurement, 
assuming $T_{d}\,=\,260\,K$, and blackbody emission from
large grains (i.e. analogous to the situation for the solar
system zodiacal dust disk). An upper limit on the fraction
of grain thermal emission to stellar emission ($f_d$\,=\,$L_{dust}/L_*$) 
was also calculated for each star using these assumptions.
While the MIRAC photometry is capable of 
detecting $\sim$4000 ${\cal Z}$ disks at T = 260\,K 
(equivalent $log(f_d)\,<\,-2.9$) around
the Tuc-Hor members observed, no convincing
mid-IR excesses were detected.

For completeness, we note that the fraction of disk luminosity
to stellar luminosity ($f_d$ = $L_{d}/L_*$) has been observed
to fall off as $f_d\,\propto\,{age}^{-1.76}$ \citep{Spangler01}. 
The disk fractional luminosity is predicted to follow 
$f_d\,\propto\,{age}^{-2}$ if the observed amount of dust
is proportional to the collision frequency of large particles,
and P-R drag is the dominant dust removal mechanism.
Recently, \citet{Dominik03} argue that for the very luminous
debris disks that have been detected so far, the collision 
timescales are much shorter than the P-R drag timescales,
all the way down to the blow-out grain size. 
For this collision-dominated scenario, \citet{Dominik03} predict
that the dust luminosity evolves as $f_d\,\propto\,{age}^{-1}$. 
While these models ignore effects like, e.g., gravitational
perturbations, or ejections, of dust-producing planetesimals by planets
(which likely had an enormous effect on the early evolution of the
asteroid belt in our solar system), they provide simple,
physically plausible models
with which to compare observations. 

If one takes the solar system zodiacal dust disk
($log(f_d)\,\simeq\,-7.1$, at log($age_{yr}$)=\,9.66), and
scale it backward in time according to Spangler et al's relation
($f_d\,\propto\,{age}^{-1.76}$) or the theoretical 
P-R drag-dominated evolution ($f_d\,\propto\,{age}^{-2}$), 
one would predict at age 30\,Myr
zodiacal dust disks with $log(f_d)\,\simeq\,-3.3$
(6900\,${\cal Z}$) or $log(f_d)\,\simeq\,-2.7$
(23000\,${\cal Z}$), respectively. Hence, for the simple
model of collisionally replenished, P-R-drag-depleted disks, 
{\it we should have easily detected the solar system's zodiacal 
dust disk at age 30\,Myr}. For the empirical relation \citep{Spangler01},
we could have detected the Sun's zody disk around most (13/15)
of the $\sim$30-Myr-old \object[T-HA]{Tuc-Hor} stars in our sample.
Backward extrapolation of the Sun's zodiacal dust disk
luminosity using Dominik \& Decin's relation for 
collisionally-dominated disks would yield  $log(f_d)\,\simeq\,-4.9$
(150\,${\cal Z}$). Such a disk would not have been detectable in 
our survey, consistent with our null result. 
{\it If} analogs of the Sun's zodiacal dust disk are
common around 30-Myr-old stars, $f_d$ must evolve as a shallower
power law ($<$1.65) than either Spangler et al.'s empirical relation
or the P-R-drag-dominated dust depletion model. 

\section{DISCUSSION \label{discussion}}

\begin{figure}
\epsscale{1.0}
\plotone{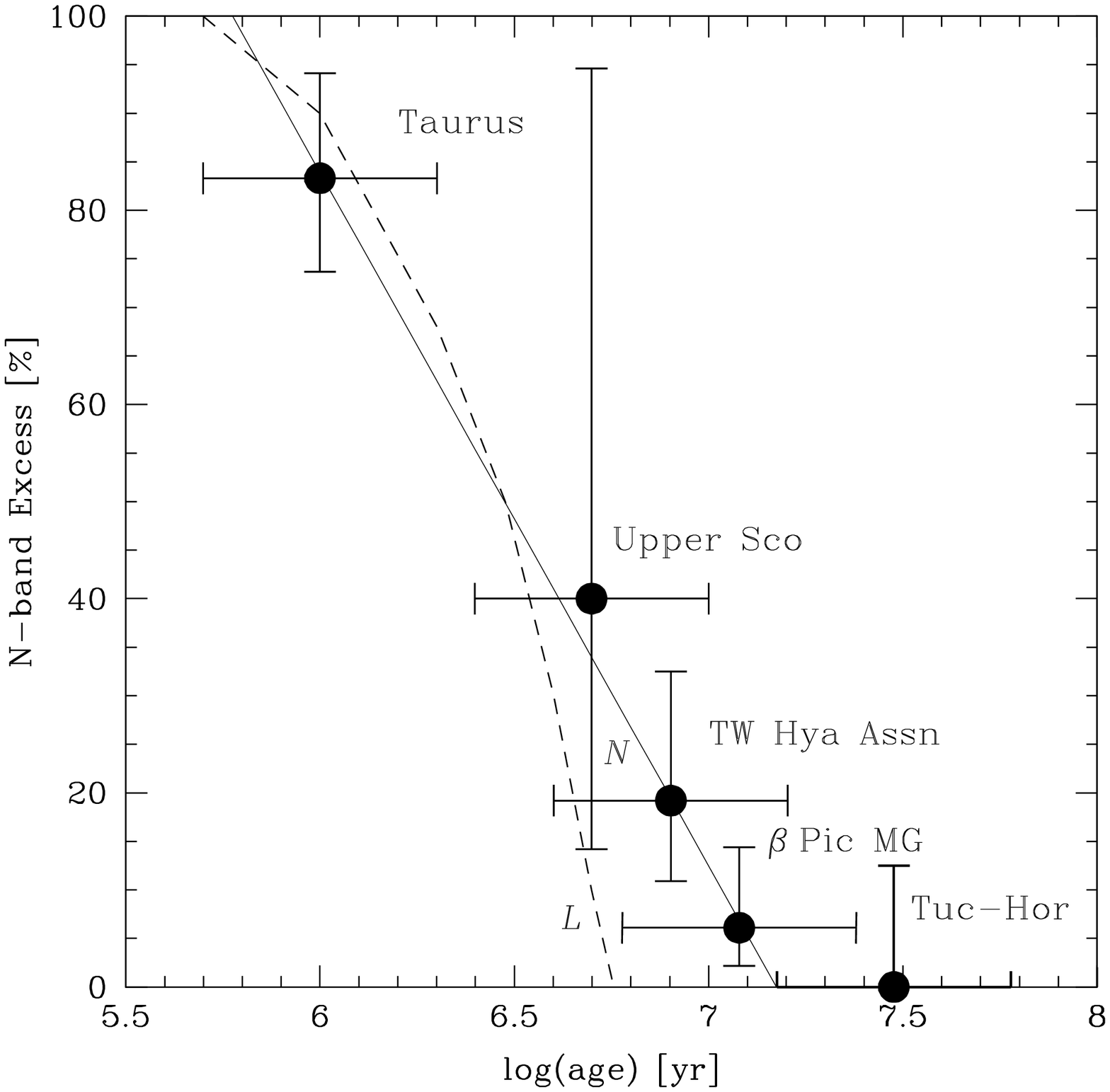}
\caption{\label{fig:xn_age}
$N$-band excess versus age for stellar samples of varying ages.
Data are plotted for the following samples:
Taurus-Auriga \citep{Kenyon95}, TW Hya Association
\citep{Jayawardhana99,Weinberger03a}, $\beta$ Pic group 
\citep{Weinberger03b}, Upper Sco, and Tuc-Hor
(both this study). Data from the IRAS 
study of AFGK-type field stars by \citet{Aumann91} 
shows that $N$-band excesses among mostly older ($\gtrsim$100 Myr) 
field stars are extraordinarily rare ($\sim$0.2\%).
{\it Dashed line} is the L-band disk fraction measured by 
\citet{Haisch01a}. The {\it solid line} is a fit 
to the $N$-band disk fraction for the four youngest groups.
It appears that $N$-band excesses are only detectable for
timescales marginally longer than that of L-band excesses,
however the statistics are still poor. Note that in the
TW Hya Association, there exists a mix of optically-thick and thin
disks, while in the $\beta$ Pic group, all of the known disks
are optically-thin. 
}
\end{figure}

\notetoeditor{Figure \ref{fig:xn_age} should go around here}

In Fig. \ref{fig:xn_age}, we plot the incidence of 
$N$-band excess versus stellar age for samples of low-mass stars. 
While $\sim$80\% of $\sim$1-Myr-old stars have in Taurus 
have $N$-band excesses \citep{Kenyon95}, only $\sim$10\% of
$\sim$10-Myr-old stars in the \object[TWA]{TW Hya Association}
and \object[bPMG]{$\beta$ Pic Moving Group} show comparable excess emission 
\citep{Jayawardhana99,Weinberger03a,Weinberger03b}. 
Of these $\sim$10-Myr-old stars, only a few are known to 
have optically-thick disks (\object{TW Hya}, \object{Hen 3-600}), 
while the others are optically-thin disks. 
Our survey imaged a small number (N\,=\,5)
of \object{Upper Sco} members ($\sim$5-Myr-old) as well, among which 
two have clear $N$-band excesses. By
an age of $\sim$30 Myr, we find that $N$-band excesses due to
optically-thick {\it or} optically-thick disks, are
rare ($\lesssim$7\%). 
Excluding the $\sim$10-Myr-old star \object[bet Pic]{$\beta$ Pic},
\citet{Aumann91} find only one star (\object[zet Lep]{$\zeta$ Lep}) 
among a sample of 548 field AFGK stars ($\approx$0.2\%)
with a convincing 12\micron\ excess.
These results seem to imply that 
{\it dust appears to be efficiently removed from $<$5-10 AU
of young stars on timescales similar to that
of the cessation of accretion.} While accretion terminates 
over a wide range of ages 
\citep[$\sim$1-10 Myr;][]{Haisch01a,Hillenbrand04}, 
the {\it duration} of the transition
from optically-thick to optically-thin has been 
observed to be remarkably short 
\citep[$\sim$10$^5$ yr;][]{Skrutskie90,Wolk96}. 

Gravitational perturbations of small planetesimals
($\sim$0.1-100\,km radius) by growing planetary
embryos ($\sim$2000\,km radius) can theoretically
cause collisional cascades of dust grains that 
produce observable mid-IR signatures
\citep{Kenyon04}. Simulations show that
when the largest planetary embryos reach radii
of $\sim$3000\,km, 
the population of dust-producing $\sim$0.1-100\,km-size 
planetesimals in the planet-forming zone becomes
collisionally depleted, and $N$-band excesses become 
undetectable. The timescale over which the $N$-band excess
would be detectable during this phase of terrestrial
planet formation is of order $\sim$1\,Myr. With
a larger sample size (e.g. FEPS Spitzer Legacy survey), 
one might be able to probe whether this is occurring around 
stars at age $\sim$30\,Myr. With 10\% 
of $\sim$10\,Myr-old stars having detectable $N$-band 
excesses \citep{Jayawardhana99,Weinberger03a,Weinberger03b},
we may be witnessing the signature from 
runaway protoplanet growth in the terrestrial planet zone
\citep{Kenyon04}.
This would agree with the isotopic evidence in our
own solar system that Moon- to Mars-sized protoplanetary
embryos accreted within the first $\sim$10-20\,Myr
\citep{Kleine02,Kleine03}, ultimately leading to the
formation of the Earth-Moon system.

\section{CONCLUSIONS \label{conclusions}}

We have undertaken a mid-IR survey of 14 young stars in the 
nearby $\sim$30-Myr-old \object[T-HA]{Tuc-Hor Association} in order 
to search for emission from warm circumstellar disks. 
No excess emission at 10\,\micron\ was detected 
around any \object[T-HA]{Tuc-Hor} members. 
If optically-thick disks do exist around these stars, 
their inner holes must be large
(range: 0.2-5.8\,AU). 
Combining our photometric results with optically-thin
dust disk models, we place the following
physical constraints on dust orbiting within $\sim$10\,AU of 
these $\sim$30-Myr-old stars: fractional disk luminosities
of $L_{dust}/L_*$ $<$ 10$^{-2.9}$ and 
dust emitting surface areas $<$4000$\times$ that of the inner 
solar system zodiacal dust. The disk masses
of micron-size dust grains with orbital radii between
the silicate dust sublimation point and $\sim$10\,AU
must be less than $\sim 10^{-6}\,M_{\oplus}$.
The photometric upper limits
also suggest that the upper limit on the
surface density of micron-sized grains is 
$\sim$10$^{-7}$ g\,cm$^{-2}$. These results imply that
inner disks dissipate on timescales comparable to the
cessation of accretion. 

\acknowledgments

For their technical support during the MIRAC runs at Magellan, 
we would like to thank Brian Duffy (Steward Observatory) and 
the staff of the Las Campanas Observatory: Emilio Cerda, Oscar 
Dulhalde, Patricio Jones, Gabriel Martin, Mauricio Navarette, 
Hernan Nu\~nez, Frank Perez, Hugo Rivera, Felipe Sanchez, Skip 
Schaller, and Geraldo Valladares. We thank Lissa Miller, Nick Siegler,
and Lynne Hillenbrand for critiquing the manuscript.
E.~E.~M. gratefully acknowledges a NASA Graduate Student 
Fellowship (NGT5-50400) for support. E.~E.~M. and M.~R.~M. are 
supported by the Spitzer Space Telescope Legacy Science Program, 
provided by NASA through Contract Number 1224768 administered by 
the Jet Propulsion Laboratory, California Institute of Technology 
under NASA contract 1407.
M.C. thanks NASA for supporting his participation in this work
through JPL contract \#1224634 with UC Berkeley.
MIRAC is supported through SAO and NSF grant AST 96-18850. 
BLINC is supported through the NASA Navigator program.

\appendix

\section{Appendix: MIRAC Photometric Calibration \label{apx_filters}}

With a significant body of \mirac\ observations acquired 
during the 2001-2003 observing runs at Magellan I, it was decided 
to calculate the photometric attributes of commonly used MIRAC 
bands on the Cohen-Walker-Witteborn (CWW) system of absolute infrared 
calibration 
\citep[e.g. ][ and references therein]{Cohen03b}. 
The photometric standard system for previously published MIRAC
studies is given in Appendix 2 of the MIRAC3 
User's Manual \citep{Hoffmann99}.

\begin{figure}
\epsscale{1.0}
\plotone{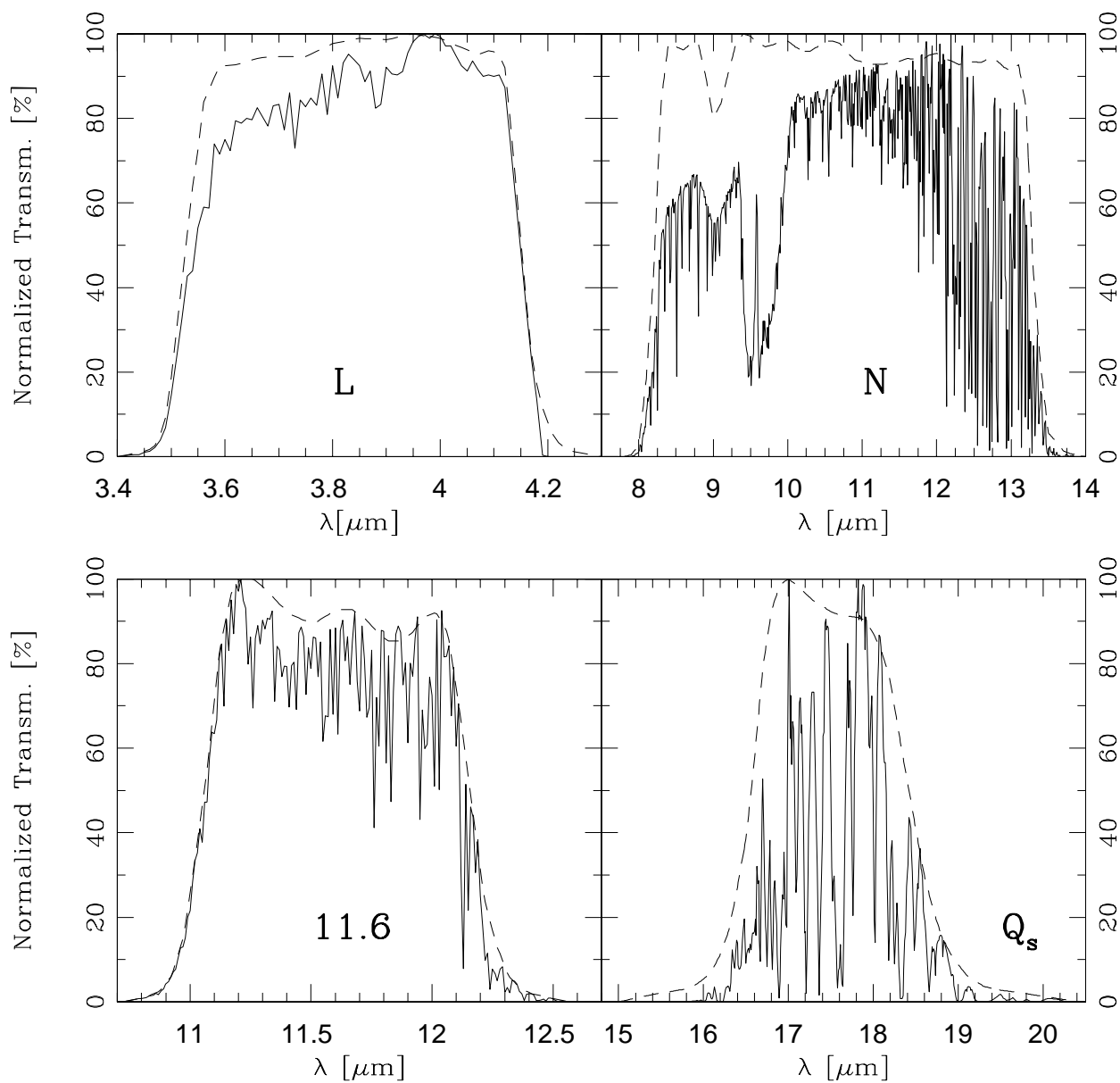}
\caption{\label{fig:filters}
Transmission profiles for the MIRAC $L$, $N$, \eleven, and \qs\ filters.
{\it Dashed lines} are the normalized filter transmission profiles. 
{\it Solid lines} are the relative spectral response curves 
(= ``RSRs'', details given in Appendix A), and represent the
product of transmissions for the filters, detector, optics,
KRS-5 dewar window, and atmosphere. 
The RSRs were used for absolutely calibrating MIRAC on the 
CWW system.}
\end{figure}

Relative spectral responses (RSRs) for each combination of
filter and window were constructed. The throughput chain consists 
of the following groups of components: atmosphere, telescope 
optics, BLINC optics, MIRAC optics, and MIRAC detector. The 
complete throughput equation
consists of the following components multiplied together: 
atmosphere, 3 aluminum mirrors (Magellan), dewar window (KRS-5 or 
KBr), KBr lens (in BLINC), 5 gold mirrors (3 in BLINC, 2 in MIRAC), 
filter, and the Si:As array. 
Most of the MIRAC observations were taken in just four of the 
seventeen filters currently available in the three MIRAC filter 
wheels: $L$, $N$, \eleven, and $Q_s$, and these were the filters 
we absolutely calibrated. 
For each MIRAC filter, we list a manufacturer's name,
mean filter wavelength ($\lambda_o$),
bandwidth ($\Delta\lambda/\lambda$; defined as the FWHM of the 
normalized transmission curve divided by the mean wavelength),
and the temperature at which the filter profile was measured 
(or extrapolated). The transmission profiles for the filters 
are plotted in Fig. \ref{fig:filters}. While we
list {\it mean} filter wavelengths ($\lambda_o$) in this discussion, 
the {\it isophotal} wavelengths ($\lambda_{iso}$) 
are given in Table \ref{zeromag}. 

\notetoeditor{Fig. \ref{fig:filters} should go around here}

The $L$ filter (OCLI ``Astro L''; $\lambda_o$ = 3.84\micron ; 
$\Delta\lambda/\lambda$ = 16.2\%; 77\,K) 
is the same one used in all previous and current MIRAC L-band 
observations,
and the transmission curve is plotted in Fig. A2.2 of \citet{Hora91}.
The $N$ filter (OCLI code W10773-8; $\lambda_o$ = 10.75\,\micron; 
$\Delta\lambda/\lambda$ = 47.2\%; Ambient) was purchased in 
1994 in preparation for comet Shoemaker-Levy-9 observations, and 
has been in use ever since. Pre-1994 MIRAC observations employed 
a slightly
bluer wideband $N$ filter whose characteristics we only present here 
for completeness (OCLI code W10575-9; $\lambda_o$ = 10.58\,\micron; 
$\Delta\lambda/\lambda$ = 45.8\%; Ambient). 
The narrow 11.6\,\micron\ filter (OCLI ``Astronomy R''; 
$\lambda_o$ = 11.62\,\micron; $\Delta\lambda/\lambda$ = 9.5\%; 
extrapolated to 5\,K) 
has been used since MIRAC was commissioned \citep{Hora91}.
Its transmission curve includes the effects of a BaF$_2$ blocker, and
it is the only filter of the four that we were able to linearly
extrapolate its transmission characteristics to the detector's 
operating temperature (5\,K; data at Ambient and 77\,K were 
available). The \qs\ or ``Q-short'' filter has also been used for 
the lifetime of MIRAC, and its characteristics are only currently 
known at ambient temperature: $\lambda_o\,=\,17.50$\,\micron; 
$\Delta\lambda/\lambda\,=\,10.6$\%).

We followed \citet{Cohen99} in using PLEXUS
\citep{Clark96} to assess mean, site-specific atmospheric transmission.
Transmission curves for KRS-5 and KBr were taken from the 
Infrared Handbook \citep{Wolfe85}. We used a KRS-5 window during 
the August 2001 and May 2002 runs, and a KBr window for the 
August 2002 and March 2003 runs. The reflectivities of the gold 
and aluminum mirrors were assumed to be flat in the wavelength 
range of interest (2-20\,\micron). The quantum efficiency 
for the MIRAC doped-silicon blocked-impurity-band (BIB) array 
was taken from \citet{Stapelbroek95}, following \citet{Hoffmann98}.
The zero magnitude attributes of the MIRAC filter systems
are given in Table \ref{zeromag}. Standard star fluxes on the
CWW system for the four primary MIRAC filters 
(with the KRS-5 dewar window) are given in Table \ref{stancal}.
When the KBr dewar window is used on \mirac, the standard star fluxes
are nearly identical (to within $<$7\% of the quoted flux 
uncertainties), so the same fluxes and magnitudes can be safely 
adopted. The flux densities in Tables \ref{tuc_phot} and
\ref{others_phot} are referenced to this system.

\section{Appendix: Comments on Individual Sources}

The only young stars in our survey to show a significant $N$-band 
excess were the T Tauri stars \object{HD 143006} and 
\object{[PZ99] J161411.0-230536}, 
both members of the $\sim$5-Myr-old \object{Upper Sco} OB subgroup. 
Both stars are targets in the FEPS Spitzer Legacy Science program,
but only \object{HD 143006} was previously known to possess a 
circumstellar disk.
Both were detected by IRAS, and the MIRAC $N$-band fluxes are 
consistent with the color-corrected 12\,\micron\, measurements in the IRAS 
FSC \citep{Moshir90}. Here we discuss these stars in more detail.

\subsection{[PZ99] J161411.0-230536 \label{PZ99}}

\object[PZ99 J161411.0-230536]{J161411} is a K0-type weak-lined T 
Tauri (EW(H$\alpha$) = 0.96\AA)
star discovered by \citet{Preibisch98}. In a spectroscopic survey to 
identify new members of \object{Upper Sco} (Mamajek, Meyer, \& Liebert, 
in prep.), the authors obtained a red, low-resolution spectrum 
of \object[PZ99 J161411.0-230536]{J161411}  in July 2000, which shows an
asymmetric H$\alpha$ feature with
blueshifted emission and redshifted absorption
(net EW(H$\alpha$) = 0.36\AA).
We confirm the strong lithium absorption 
(EW(Li $\lambda$6707) = 0.45\,\AA) observed by
\citet{Preibisch98}. The UCAC2 proper motion 
\citep{Zacharias03} for \object[PZ99 J161411.0-230536]{J161411} 
is consistent with membership
in the \object{Upper Sco} subgroup. The \kn\, color (=\,2.1) is similar
to that of classical T Tauri stars in \object{Taurus-Auriga}
\citep{Kenyon95}. The photometric 
and spectroscopic evidence suggest that this $\sim$5-Myr-old,
$\sim$1\,M$_\odot$ star is actively accreting from a circumstellar disk.

\subsection{HD 143006 \label{HD143006}} 

\object{HD 143006} is a G5Ve \citep{Henize76} T Tauri star 
with strong Li absorption \citep[EW(Li\,$\lambda$6707) 
= 0.24\AA;][]{Dunkin97}.
The star is situated in the middle of the \object{Upper
Sco} OB association, and its proper motion 
\citep[$\mu_\alpha$,\,$\mu_\delta$ = -11, 
-20 mas/yr;][]{Zacharias03} and radial velocity 
\citep[-0.9\,km/s;][]{Dunkin97} are indistinguishable
from other association members \citep{deBruijne99}.
Several studies have classified HD 143006
as a distant G-type supergiant or ``pre-planetary
nebula'' \citep{Carballo92,Kohoutek01}, however
we believe this is erroneous. 
If \object{HD 143006} were indeed a supergiant at $d$ = 3.4\,kpc
\citep{Pottasch88}, its tangential velocity
would be $\sim$370 km/s -- extraordinarily fast for 
a population I star. The MIRAC $N$ and \eleven\ photometry
agrees well with the data points in the SED for
\object{HD 143006} plotted in Fig. 2 of \citet{Sylvester96}. 
The spectral energy distribution for
\object{HD 143006} and its optically-thick disk
is well-studied from 0.4-1300\,\micron, so
we do not discuss this object further.




\clearpage
\begin{deluxetable}{llcll}
\tabletypesize{\scriptsize}
\setlength{\tabcolsep}{0.03in} 
\tablewidth{0pt}
\tablecaption{Age Estimates for Tucana-Horologium Association \label{tuc_age}}
\tablehead{
{(1)}       & {(2)}    & {(3)}   & {(4)}    & {(5)} \\
{Reference} & {Group}  & {Age}   & {Method} & {Notes}\\
{\ldots}    & {\ldots} & {(Myr)} & {\ldots} & {\ldots}
}
\startdata
\citet{Zuckerman00} & Tuc     & 40    & H$\alpha$ Emission     & Comparing H$\alpha$ emission of 3 stars to $\alpha$ Per members\\ 
\citet{Torres00}    & Hor     & 30    & Theoretical Isochrones & \citet{Siess97} tracks \\
\citet{Torres01}    & Tuc-Hor & 20    & Velocity Dispersion    & ``GAYA'' = Tuc-Hor \\
\citet{Stelzer01}   & Tuc     & 10-30 & X-ray Emission         & Member L$_X$ values similar to TWA, Tau-Aur, \& IC 2602\\
\citet{Zuckerman01a}& Tuc     & 10-40 & Theoretical Isochrones & K \& M stars; \citet{Siess00} tracks\\ 
\enddata
\end{deluxetable}

\clearpage
\begin{deluxetable}{llccl}
\tabletypesize{\scriptsize}
\setlength{\tabcolsep}{0.03in} 
\tablewidth{0pt}
\tablecaption{MIRAC Observations of Tuc-Hor Members \label{tuc_obs}}
\tablehead{
{(1)}  & {(2)}  & {(3)}    & {(4)}       & {(5)}  \\
{UT}   & {Star} & {Band}   & {On-Source} & {Flux}  \\ 
{Date} & {Name} & {\ldots} & {Time (s)}  & {Standards}  
}
\startdata
2001 Aug 8  & HIP 105388    & $N$ & 300 & $\iota$ Cet, $\alpha$ CMa\\
\nodata     & HIP 105404    & $N$ & 600 & $\iota$ Cet, $\alpha$ CMa\\
\nodata     & HIP 107947    & $N$ & 600 & $\iota$ Cet, $\alpha$ CMa\\
\nodata     & HIP 108195    & $N$ & 360 & $\iota$ Cet, $\alpha$ CMa\\
\nodata     & HIP 116748AB  & $N$ & 600 & $\iota$ Cet, $\alpha$ CMa\\
\nodata     & HIP 1481      & $N$ & 480 & $\iota$ Cet, $\alpha$ CMa\\
\nodata     & HIP 1910      & $N$ & 840 & $\iota$ Cet, $\alpha$ CMa\\
\nodata     & HIP 2729      & $N$ & 840 & $\iota$ Cet, $\alpha$ CMa\\
2001 Aug 9  & HIP 490       & $N$ & 420 & $\iota$ Cet, $\eta$ Sgr, $\alpha$ CMa\\
\nodata     & HIP 6485      & $N$ & 600 & $\iota$ Cet, $\eta$ Sgr, $\alpha$ CMa\\
\nodata     & HIP 6856      & $N$ & 660 & $\iota$ Cet, $\eta$ Sgr, $\alpha$ CMa\\
\nodata     & HIP 9892      & $N$ & 840 & $\iota$ Cet, $\eta$ Sgr, $\alpha$ CMa\\
\nodata     & ERX 37N       & $N$ &1020 & $\iota$ Cet, $\eta$ Sgr, $\alpha$ CMa\\
2001 Aug 10 & HIP 9685      & $N$ & 690 & $\alpha$ Cen, $\iota$ Cet, $\eta$ Sgr\\
2002 Aug 21 & HIP 1910      & $N$ & 180 & $\gamma$ Cru, $\eta$ Sgr, $\iota$ Cet\\
\enddata
\end{deluxetable}

\clearpage
\begin{deluxetable}{llccl}
\tabletypesize{\scriptsize}
\setlength{\tabcolsep}{0.03in} 
\tablewidth{0pt}
 \tablecaption{MIRAC Observations of Other Stars \label{others_obs}}
\tablehead{
{(1)}  & {(2)}  & {(3)}    & {(4)}       &  {(5)}\\
{UT }  & {Star} & {Band}   & {On-Source} &  {Flux}\\ 
{Date} & {Name} & {\ldots} & {Time (s)}  &  {Standards}
}
\startdata
2001 Aug 6 & GJ 799 AB       & 11.6     & 135    & $\iota$ Cet \\
\nodata    & GJ 803          & 11.6     & 165    & $\iota$ Cet \\
\nodata    & HIP 108195      & 11.6     & 60     & $\iota$ Cet \\
2001 Aug 7 & HIP 99273       & 11.6     & 255    & $\alpha$ Car, $\alpha$ PsA, $\gamma$ Cru \\
2001 Aug 8 & HD 143006       & $N$        & 60     & $\iota$ Cet, $\alpha$ CMa \\
\nodata    & HD 143006       & 11.6     & 120    & $\gamma$ Cru\\
\nodata    & $[PZ99]$ J161318.6-221248 & $N$ & 600 & $\iota$ Cet, $\alpha$ CMa \\
\nodata    & HD 181327       & $N$        & 240    & $\iota$ Cet, $\alpha$ CMa\\
\nodata    & HR 7329         & $N$        &  60    & $\iota$ Cet, $\alpha$ CMa\\
2001 Aug 9 & RX J1853.1-3609 & $N$        & 960    & $\iota$ Cet, $\eta$ Sgr, $\alpha$ CMa\\
\nodata    & RX J1917.4-3756 & $N$        & 660    & $\iota$ Cet, $\eta$ Sgr, $\alpha$ CMa\\
2001 Aug 10& HIP 63797       & $N$        & 300    & $\alpha$ Cen, $\iota$ Cet, $\eta$ Sgr\\
\nodata    & $[PZ99]$ J161411.0-230536 & $N$ & 360 & $\alpha$ Cen, $\iota$ Cet, $\eta$ Sgr\\
\nodata    & ScoPMS 214      & $N$        & 450    & $\alpha$ Cen, $\iota$ Cet, $\eta$ Sgr\\
\nodata    & ScoPMS 5        & $N$        & 450    & $\alpha$ Cen, $\iota$ Cet, $\eta$ Sgr\\ 
\nodata    & HIP 95149       & $N$        & 360    & $\alpha$ Cen, $\iota$ Cet, $\eta$ Sgr\\ 
\nodata    & HIP 113579      & $N$        & 510    & $\alpha$ Cen, $\iota$ Cet, $\eta$ Sgr\\ 
\nodata    & HIP 1134        & $N$        & 480    & $\alpha$ Cen, $\iota$ Cet, $\eta$ Sgr\\ 
2002 Aug 21& HIP 93815       & $N$        & 180    & $\gamma$ Cru, $\eta$ Sgr, $\iota$ Cet\\
\nodata    & HIP 99803 A     & $N$        & 690    & $\gamma$ Cru, $\eta$ Sgr, $\iota$ Cet\\
\nodata    & HIP 99803 B     & $N$        & 165    & $\gamma$ Cru, $\eta$ Sgr, $\iota$ Cet\\
\nodata    & HIP 105441      & $N$        & 360    & $\gamma$ Cru, $\eta$ Sgr, $\iota$ Cet\\
\nodata    & HIP 107649      & $N$        & 240    & $\gamma$ Cru, $\eta$ Sgr, $\iota$ Cet\\
\nodata    & PPM 366328      & $N$        & 180    & $\gamma$ Cru, $\eta$ Sgr, $\iota$ Cet\\
\nodata    & HIP 108809      & $N$        & 360    & $\gamma$ Cru, $\eta$ Sgr, $\iota$ Cet\\
\nodata    & HIP 108422      & $N$        & 300    & $\gamma$ Cru, $\eta$ Sgr, $\iota$ Cet\\
\nodata    & GJ 879          & $N$        & 240    & $\gamma$ Cru, $\eta$ Sgr, $\iota$ Cet\\
\enddata 
\end{deluxetable}

\clearpage
\begin{deluxetable}{lllccccrc}
\tabletypesize{\scriptsize}
\setlength{\tabcolsep}{0.03in} 
\tablewidth{0pt}
\tablecaption{$N$-band Photometry of Tuc-Hor Members \label{tuc_phot}}
\tablehead{
{(1)}       &{(2)}       & {(3)}   & {(4)}   & {(5)}            & {(6)}            & {(7)}    & {(8)}\\
{Name}      &{Name}      & {Spec.} & {\ks}   & {Pred.\,F$_\nu$} & {Meas.\,F$_\nu$} & {E(N)}   & {Dev.}\\ 
{\ldots}    &{\ldots}    & {Type}  & {(mag)} & {(mJy)}          & {(mJy)}          & {(mJy)}  & {($\sigma$)}}
\startdata
HIP 490     & HD 105     & G0 & 6.12\,$\pm$\,0.02 & 139 &138\,$\pm$\,8~                  &  ~--1\,$\pm$\,8~ & --0.2\\        
HIP 1481    & HD 1466    & F8 & 6.15\,$\pm$\,0.02 & 135 &151\,$\pm$\,20                  &  +16\,$\pm$\,20 & +0.8\\ 
HIP 1910    & BPM 1699   & M0 & 7.49\,$\pm$\,0.02 & ~51 &~49\,$\pm$\,7~                  &  ~--2\,$\pm$\,7~ & --0.3\\   
HIP 2729    & HD 3221    & K4 & 6.53\,$\pm$\,0.02 & 111 &102\,$\pm$\,5~                  &  ~--9\,$\pm$\,5~ & --1.7\\     
HIP 6485    & HD 8558    & G6 & 6.85\,$\pm$\,0.03 & ~71 &~82\,$\pm$\,4~                  &  +11\,$\pm$\,4~ & +2.4\\    
HIP 6856    & HD 9054    & K2 & 6.83\,$\pm$\,0.02 & ~72 &~85\,$\pm$\,4~                  &  +13\,$\pm$\,4~ & +3.0\\    
HIP 9685    & HD 12894   & F4 & 5.45\,$\pm$\,0.02 & 258 &207\,$\pm$\,17                  &  --51\,$\pm$\,18 & --2.9\\ 
HIP 9892    & HD 13183   & G5 & 6.89\,$\pm$\,0.02 & ~68 &~67\,$\pm$\,6~                  &  ~--1\,$\pm$\,6~ & --0.2\\   
ERX 37N     & AF Hor     & M3 & 7.64\,$\pm$\,0.03 & ~42 &~46\,$\pm$\,3~                  &  ~+4\,$\pm$\,3~ & +1.3\\  
HIP 105388  & HD 202917  & G5 & 6.91\,$\pm$\,0.02 & ~67 &~65\,$\pm$\,9~                  &  ~--2\,$\pm$\,9~ & --0.3\\        
HIP 105404  & HD 202947  & K0 & 6.57\,$\pm$\,0.02 & ~92 &~74\,$\pm$\,9~                  &  --18\,$\pm$\,9~ & --1.9\\        
HIP 107947  & HD 207575  & F6 & 6.03\,$\pm$\,0.02 & 151 &155\,$\pm$\,12                  &  ~+4\,$\pm$\,12 & +0.3\\
HIP 108195  & HD 207964  & F3 & 4.91\,$\pm$\,0.02 & 424 &394\,$\pm$\,20\tablenotemark{a} &  --30\,$\pm$\,21 & --1.4\\
HIP 116748A & HD 222259A & G6 & 6.68\,$\pm$\,0.03 & ~83 &~75\,$\pm$\,12                  &  ~--8\,$\pm$\,12 & --0.7\\
HIP 116748B & HD 222259B & K~ & 7.03\,$\pm$\,0.06 & ~60 &~53\,$\pm$\,14                  &  ~--7\,$\pm$\,14 & --0.5\\   
\enddata
\tablecomments{Columns 
(1) Hipparcos name, 
(2) other name, 
(3) spectral type, from either ZW00, TDQ00, or ZSW01,
(4) \ks\ magnitude from 2MASS \citep{Cutri03},
(5) measured MIRAC $N$-band flux,
(6) predicted photospheric flux,
(7) flux excess and uncertainty,
(8) residual deviation = $E(N)/\sigma(E(N))$.
ERX 37N is given in SIMBAD as [TDQ2000] ERX 37N.
Predicted $N$-band photospheric fluxes use or assume: 
2MASS \ks\ magnitudes, \av\ = 0, dwarf color relations from \S\ref{results}, 
and zero magnitude flux of 37.25 Jy for $N$-band.
}
\tablenotetext{a}{HIP 108195 was also imaged at 11.6\micron~ with a flux of 284\,$\pm$\,40\,mJy.}
\end{deluxetable}

\clearpage
\begin{deluxetable}{lllcccccr}
\tabletypesize{\scriptsize}
\setlength{\tabcolsep}{0.03in} 
\tablewidth{0pt}
\tablecaption{Measured Photometry for Other Stars \label{others_phot}}
\tablehead{
 {(1)}         &{(2)}    &{(3)}   & {(4)}   & {(5)}    & {(6)}            & {(7)}            &  {(8)}      & {(9)}\\
 {Name}        &{Name}   &{Spec.} & {K$_s$} & {Band}   & {Pred.\,F$_\nu$} & {Meas.\,F$_\nu$} &  {E(N)}     & {Dev.}\\ 
 {\ldots}      &{\ldots} &{Type}  & {(mag)} & {\ldots} & {(mJy)}          & {(mJy)}          &  {(mJy)}    & {($\sigma$)} \\
}			  				  			 	   
\startdata
\cutinhead{Tuc-Hor Rejects}
HD 177171                    & HIP 93815        & F7    & 3.81\,$\pm$\,0.10\tablenotemark{a} & $N$    & 1171 & 1236\,$\pm$\,103 &  ~+65\,$\pm$\,149 & ~+0.4\\
HD 191869 A\tablenotemark{b} & HIP 99803 A      & F7    & 6.81\,$\pm$\,0.02                  & $N$    & ~~74 & ~~87\,$\pm$\,9~~ &  ~+13\,$\pm$\,9~~ & ~+1.5\\
HD 191869 B\tablenotemark{b} & HIP 99803 B    & \nodata & 6.86\,$\pm$\,0.03                  & $N$    & ~~70 & ~~63\,$\pm$\,14~ & ~~--7\,$\pm$\,14~ &~--0.5\\
HD 202746                    & HIP 105441       & K2    & 6.40\,$\pm$\,0.02                  & $N$    & ~107 & ~116\,$\pm$\,11~ &  ~~+9\,$\pm$\,11~ & ~+0.8\\
HD 207129                    & HIP 107649       & G0    & 4.12\,$\pm$\,0.02\tablenotemark{c} & $N$    & ~880 & ~837\,$\pm$\,70~ & ~--43\,$\pm$\,72~ &~--0.6\\
PPM 366328                   & TYC 9129-1361-1  & K0    & 7.61\,$\pm$\,0.02                  & $N$    & ~~35 & ~~63\,$\pm$\,24~ &  ~+28\,$\pm$\,24~ & ~+1.1\\
HD 208233\tablenotemark{d}   & HIP 108422       & G8    & 6.75\,$\pm$\,0.02                  & $N$    & ~~78 & ~~89\,$\pm$\,14~ &  ~+11\,$\pm$\,14~ & ~+0.8\\
\cutinhead{Upper Sco Members + FEPS }
$[PZ99]$ J161411.0-230536    & TYC 6793-819-1   & K0    & 7.46\,$\pm$\,0.03                  & $N$    & ~~48 & ~273\,$\pm$\,11~ &  +225\,$\pm$\,11~ & +20.3\\
ScoPMS 214                   & NTTS 162649-2145~& K0    & 7.76\,$\pm$\,0.02                  & $N$    & ~~42 & ~~37\,$\pm$\,5~~ & ~~--5\,$\pm$\,5~~ &~--1.0\\
ScoPMS 5                     & HD 142361        & G3    & 7.03\,$\pm$\,0.02\tablenotemark{e} & $N$    & ~~60 & ~~58\,$\pm$\,6~~ & ~~--2\,$\pm$\,6~~ &~--0.4\\
HD 143006                    & HBC 608          & G6/8~ & 7.05\,$\pm$\,0.03                  & $N$    & ~134 & ~648\,$\pm$\,31~ &  +514\,$\pm$\,31~ & +16.5\\
HD 143006                    & HBC 608          & G6/8~ & 7.05\,$\pm$\,0.03                  & 11.6 & ~104 & ~640\,$\pm$\,34~ &  +536\,$\pm$\,34~ & +15.7\\
$[PZ99]$ J161318.6-221248    & TYC 6213-306-1   & G9    & 7.43\,$\pm$\,0.02                  & $N$    & ~~45 & ~~32\,$\pm$\,6~~ & ~--13\,$\pm$\,6~~ &~--2.2\\
\cutinhead{$\beta$ Pic Group Members}	  			   			   
GJ 799 A                     & HIP 102141 A     & M4.5~ & 5.70\,$\pm$\,0.10\tablenotemark{f} & 11.6 & ~202 & ~259\,$\pm$\,24~ &  ~+57\,$\pm$\,30~ & ~+1.9\\
GJ 799 B                     & HIP 102141 B     & M4    & 5.70\,$\pm$\,0.10\tablenotemark{f} & 11.6 & ~202 & ~260\,$\pm$\,25~ &  ~+58\,$\pm$\,31~ & ~+1.9\\
GJ 803                       & HIP 102409       & M0    & 4.53\,$\pm$\,0.02                  & 11.6 & ~627 & ~608\,$\pm$\,32~ &  --19\,$\pm$\,34~ &~--0.6\\
HD 181327                    & HIP 95270        & F5/6~ & 5.91\,$\pm$\,0.03                  & $N$    & ~169 & ~200\,$\pm$\,18~ &  ~+31\,$\pm$\,19~ & ~+1.7\\
HR 7329                      & HIP 95261        & A0    & 5.01\,$\pm$\,0.03                  & 11.6 & ~301 & ~343\,$\pm$\,31~ &  ~+42\,$\pm$\,32~ & ~+1.3\\
HR 7329                      & HIP 95261        & A0    & 5.01\,$\pm$\,0.03                  & $N$    & ~387 & ~466\,$\pm$\,52~ &  ~+79\,$\pm$\,53~ & ~+1.5\\
\cutinhead{CrA Off-Cloud Stars + FEPS}
RX J1853.1-3609              & HD 174656        & G6    & 7.28\,$\pm$\,0.02                  & $N$    & ~~48 & ~~46\,$\pm$\,4~~ &  ~--2\,$\pm$\,4~~ &~--0.4\\
RX J1917.4-3756              & SAO 211129       & K2    & 7.47\,$\pm$\,0.03                  & $N$    & ~~44 & ~~45\,$\pm$\,4~~ &   ~+1\,$\pm$\,4~~ & ~+0.3\\
\cutinhead{Sco-Cen Reject}
HD 113376\tablenotemark{g}   & HIP 63797        & G3    & 6.70\,$\pm$\,0.02                  & $N$    & ~~81 & ~~83\,$\pm$\,9~~ &   ~+2\,$\pm$\,9~~ & ~+0.2\\ 
\cutinhead{Other FEPS Targets}
HD 181321                    & HIP 95149        & G5    & 4.93\,$\pm$\,0.02                  & $N$    & ~418 & ~425\,$\pm$\,21~ &  ~~+7\,$\pm$\,22~ & ~+0.3\\  
HD 191089                    & HIP 99273        & F5    & 6.08\,$\pm$\,0.03                  & 11.6 & ~113 & ~132\,$\pm$\,15~ &  ~+19\,$\pm$\,15~ & ~+1.3\\
HD 209253                    & HIP 108809       & F6/7  & 5.39\,$\pm$\,0.02                  & $N$    & ~273 & ~255\,$\pm$\,21~ & ~--18\,$\pm$\,22~ &~--0.8\\
HD 216803                    & GJ 879           & K4    & 3.81\,$\pm$\,0.02\tablenotemark{c} & $N$    & 1233 & 1027\,$\pm$\,85~ & --206\,$\pm$\,88~ &~--2.3\\ 
HD 217343                    & HIP 113579       & G3    & 5.94\,$\pm$\,0.03                  & $N$    & ~164 & ~160\,$\pm$\,8~~ & ~~--4\,$\pm$\,9~~ &~--0.4\\
HD 984                       & HIP 1134         & F5    & 6.07\,$\pm$\,0.02                  & $N$    & ~145 & ~131\,$\pm$\,14~ & ~--14\,$\pm$\,14~ &~--1.0\\ 
\enddata
\tablecomments{
(1) Hipparcos name, 
(2) other name, 
(3) spectral type (from SIMBAD unless otherwise noted)
(4) \ks\, magnitude from 2MASS \citep[][]{Cutri03}, unless otherwise noted,
(5) measured MIRAC $N$-band flux,
(6) predicted photospheric flux,
(7) flux excess and uncertainty,
(8) residual deviation = $E(N)/\sigma(E(N))$.
}

\tablenotetext{a}{Star is saturated in 2MASS. We adopt the V magnitude from Hipparcos
\citep{ESA97}, and the intrinsic \vj\ and \jk\ color for F7 stars from 
\citet{Kenyon95} (converted to 2MASS system via Carpenter 2001), to calculate a
rough $K_s$ magnitude. We assume an uncertainty of 0.10 mag.}
\tablenotetext{b}{\citet{Zuckerman01b} calls the pair HIP 99803 NE and SW. 
A = SE and B = NW.}
\tablenotetext{c}{2MASS photometry is saturated. We take the \kcit\
magnitude from \citet{Aumann91} and transform it to the 2MASS system
via equation (12) of \citet{Carpenter01}.}
\tablenotetext{d}{We can not rule out HIP 108422 as a Tuc-Hor 
member based on its proper motion, or the agreement between the calculated cluster parallax 
and Hipparcos trigonometric parallax. However, we conservatively exclude the star 
as a member at present, since no spectroscopic evidence of youth has
been presented in the literature. If it is co-moving with the Tucana nucleus, 
we predict a RV of +3\,\kms.}
\tablenotetext{e}{A 0.8'' binary discovered by \citet{Ghez93} and seen in MIRAC 
K-band images. MIRAC $N$ and 2MASS \ks\ magnitudes are for unresolved pair.} 
\tablenotetext{f}{The 2MASS \ks\ magnitudes from \citet{Reid02} appear
to be at odds with the combined magnitude for A \& B measured by \citet{Cutri03},
\citet{Nelson86}, and \citet{Probst83}. The system is essentially an equal brightness
binary at optical bands as well as at N, so we split the 2MASS \ks\ magnitude evenly
and adopt a generous 0.10 mag error. Spectral types are from \citet{Hawley96}}
\tablenotetext{g}{Rejected as a Sco-Cen member by \citet{Mamajek02}}.

\end{deluxetable}

\clearpage
\begin{deluxetable}{lcccccccccc}
\tabletypesize{\scriptsize}
\setlength{\tabcolsep}{0.03in} 
\tablewidth{0pt}
\tablecaption{Model Parameters for Tuc-Hor Members\label{tuc_model}}
\tablehead{           
{(1)} &{(2)}  &{(3)}      &{(4)}          &{(5)}        &{(6)}      &{(7)}             &{(8)}          &{(9)}         &{(10)}        &{(11)}\\
{Name}&{$\pi$}&{log}      &{log}          &{r$_{hole}$}&{$\bar{a}$}&{$r_{in}-r_{out}$}&{$\Sigma_o$}   &{M$_{disk}$}  &{log}          &{Zodys}\\ 
{\ldots} &{(mas)}&{T$_{eff}$}&{L/L$_{\odot}$}&{(AU)}      &{(\micron)}&{(AU)}            &{(g cm$^{-2}$)}&{(M$_\oplus$)}&{(L$_d$/L$_*$)}&{(${\cal Z}$)}\\
} 
\startdata
HIP 490    & 24.9 & 3.776 &  +0.17 & 0.3 & 0.50 & 0.05-13.9 &3.4E-06 &7.8E-05 & -3.25 & 2.6E3\\
HIP 1481   & 24.4 & 3.780 &  +0.21 & 1.8 & 0.53 & 0.06-13.8 &8.2E-06 &1.9E-04 & -2.89 & 6.4E3\\
HIP 1910   & 21.6 & 3.585 & --0.81 & 0.1 & 0.10 & 0.03-10.8 &1.4E-05 &1.9E-04 & -2.20 & 2.9E3\\
HIP 2729   & 21.8 & 3.643 & --0.32 & 0.7 & 0.27 & 0.04-12.4 &5.0E-06 &9.2E-05 & -2.80 & 2.1E3\\
HIP 6485   & 20.3 & 3.744 & --0.04 & 0.3 & 0.34 & 0.05-14.4 &3.2E-06 &7.7E-05 & -3.10 & 2.0E3\\
HIP 6856   & 26.9 & 3.707 & --0.55 & 0.1 & 0.12 & 0.04-15.3 &2.6E-06 &7.1E-05 & -2.85 & 1.1E3\\
HIP 9685   & 21.2 & 3.829 &  +0.71 & 4.9 & 1.44 & 0.09-12.0 &6.9E-06 &1.2E-04 & -3.45 & 7.5E3\\
HIP 9892   & 19.9 & 3.752 & --0.08 & 0.2 & 0.31 & 0.05-14.9 &4.5E-06 &1.2E-04 & -2.91 & 3.0E3\\
ERX 37N    & 22.6 & 3.540 & --0.89 & 0.1 & 0.14 & 0.02-8.6  &8.3E-06 &7.1E-05 & -2.48 & 1.2E3\\
HIP 105388 & 21.8 & 3.746 & --0.16 & 0.2 & 0.26 & 0.05-15.3 &5.4E-06 &1.5E-04 & -2.75 & 3.6E3\\
HIP 105404 & 21.7 & 3.719 & --0.21 & 0.2 & 0.26 & 0.04-14.6 &6.4E-06 &1.6E-04 & -2.68 & 3.8E3\\
HIP 107947 & 22.2 & 3.795 &  +0.38 & 2.1 & 0.75 & 0.06-13.0 &6.3E-06 &1.3E-04 & -3.16 & 4.8E3\\
HIP 108195 & 21.5 & 3.817 &  +0.91 & 7.9 & 2.40 & 0.11-11.7 &1.0E-05 &1.6E-04 & -3.53 & 8.6E3\\
HIP 116748A& 21.6 & 3.746 & --0.09 & 1.0 & 0.31 & 0.05-14.8 &8.0E-06 &2.1E-04 & -2.65 & 5.0E3\\
HIP 116748B& 21.6 & 3.645 & --0.46 & 0.2 & 0.19 & 0.04-13.0 &1.4E-05 &2.9E-04 & -2.24 & 5.8E3\\
\enddata
\tablecomments{
(1) Star name.
(2) Parallax. ERX 37N parallax calculated via cluster parallax method, the other values are from 
Hipparcos \citep{ESA97}.
(3) Stellar effective temperature.
(4) Luminosity in solar units. 
(5) Lower limits on inner hole radius for a hypothetical optically-thick disk (\S\ref{optthick}). 
(6) Mean calculated grain size, where $\bar{a}$ = 5$a_{min}$/3, where $a_{min}$ is the
blow-out grain size (Eqn. 1; \S\ref{optthin}). 
(7) Inner and outer radii for calculation of optically-thin disk (\S\ref{optthin}).
(8) Disk surface mass density for optically-thin disk model (\S\ref{optthin});
independent of radius for our adopted model with
($\Sigma\,=\,\Sigma_o\,r_{AU}^{-p}$; $p\,=\,0$).
(9) Upper limit on disk mass (in grains of size $\bar{a}$) for optically-thin 
model (\S\ref{optthin}).
(10) Upper limit to fractional luminosity of scaled-up zodiacal dust model (\S\ref{zody}).
(11) Upper limit to emitting area of scaled-up zodiacal dust model (in units
of ``zodys'', where 1\,${\cal Z}$ = 10$^{21}$\,cm$^2$; \S\ref{zody}).
}
\end{deluxetable}

\clearpage
\begin{deluxetable}{llllll}
\tabletypesize{\scriptsize}
\setlength{\tabcolsep}{0.03in} 
\tablewidth{0pt}
\tablecaption{Effects of Changing Adopted Values on Model Output \label{effects_of_change}}
\tablehead{
{(1)}       & {(2)}              & {(3)}               & {(4)}               & {(5)}               & {(6)}\\
{Parameter} & {$\Delta$r$_{in}$} & {$\Delta$r$_{out}$} & {$\Delta$$\Sigma_o$}& {$\Delta$$M_{dust}$}& {Notes}
}
\startdata
$\beta$ = 2       & $\times$(1.00-1.17) & $\times$(1.4-2.3) & /(1.3-0.26) & $\times$(1.5-21) & crystalline case\\
$\beta$ = 1       & /(1.00-1.16) & /(1.4-2.3) & $\times$(1.2-0.24) & /(1.6-22)   & amorphous case\\
$\beta$ = 0       & /(1.01-1.74) & /(2.7-12)  & $\times$(1.5-0.013)& /(4.7-10700)& blackbody case \\
$\bar{a}\times10$ & /(1.01-1.73) & /(1.6-4.4) & $\times$(9.5-0.32) & $\times$(3.8-0.017)& \nodata\\
$\bar{a}\times100$& /(1.01-1.75) & /(1.6-7.1) & $\times$(95-2.7)   & $\times$(38-0.054) & \nodata\\
$p$ = 0.34        & no change    & no change  & $\times$(1.3-0.6)  & /(1.5-2.8)  & zodiacal case\\
$p$ = 1           & no change    & no change  & $\times$(1.3-0.17) & /(4.7-26)   & \nodata\\
$p$ = 1.5         & no change    & no change  & /(1.1-21)   & /(13-140)   & min. mass solar nebula case\\
\enddata
\tablecomments{Columns: 
(1) model parameter, 
(2-5) range of factors acted upon model values in Table \ref{tuc_model} if
parameter in column \#1 is adopted, 
(6) case name.}
\end{deluxetable}

\clearpage
\begin{deluxetable}{lccccccc}
\tabletypesize{\scriptsize}
\setlength{\tabcolsep}{0.02in} 
\tablewidth{0pt}
\tablecaption{Zero-Magnitude Attributes of MIRAC Photometric Bands \label{zeromag}}
\tablehead{
{(1)} & {(2)} & {(3)} & {(4)} & {(5)} & {(6)} & {(7)} & {(8)}\\
{MIRAC} & {$\lambda_{iso}$} & {Bandwidth} & {In-Band} & {F$_{\lambda}$(iso)} & {Bandwidth} & {F$_{\nu}$(iso)} & {$\nu$(iso)}\\ 
{Band} & {(\micron)} & {(\micron)} & {(W\,cm$^{-2}$)} & {(W\,cm$^{-2}$\,\micron$^{-1}$)} & {(Hz)} & {(Jy)} & {(Hz)}   
}
\startdata
 $L$          & ~3.844 & 0.5423 & 2.631E-15 & 4.852E-15 & 1.102E+13 & 238.8~ & 7.793E+13 \\
~~~uncert.    & ~0.018 & 0.0037 & 1.609\%   & 8.469E-17 & 6.820E+10 & ~~4.1~ & 7.361E+11 \\
 $N$          & 10.35~ & 3.228~ & 3.263E-16 & 1.011E-16 & 8.760E+12 & ~37.25 & 2.946E+13 \\
~~~uncert.    & ~0.05~ & 0.022~ & 1.632\%   & 1.789E-18 & 4.170E+10 & ~~0.60 & 2.416E+11 \\
 \eleven      & 11.57~ & 0.8953 & 5.816E-17 & 6.496E-17 & 2.006E+12 & ~29.00 & 2.592E+13 \\
~~~uncert.    & ~0.08~ & 0.0135 & 2.110\%   & 1.686E-18 & 2.149E+10 & ~~0.61 & 2.827E+11 \\
 \qs          & 17.58~ & 0.9130 & 1.123E-17 & 1.230E-17 & 8.834E+11 & ~12.72 & 1.706E+13 \\ 
~~~uncert.    & ~0.14~ & 0.0185 & 2.494\%   & 3.951E-19 & 1.263E+10 & ~~0.32 & 2.171E+11 \\
\enddata
\tablecomments{Columns 
(1) name of MIRAC band,
(2) isophotal wavelength,
(3) wavelength bandwidth of RSR, 
(4) In-band flux for zero-magnitude star,
(5) isophotal monochromatic intensity (wavelength units),
(6) frequency bandwidth of RSR,
(7) isophotal monochromatic intensity (frequency units),
(8) isophotal frequency.
Note that the stated quantities assume that a KRS-5 dewar window is used.
If the KBr dewar window is used, the values are nearly identical.
For KBr, every stated value is within 5\% of the stated uncertainty for the $L$,
\eleven, and \qs\ bands, and within 36\% of the stated uncertainty
for $N$-band. 
}
\end{deluxetable}

\clearpage
\begin{deluxetable}{lllccccccc}
\tabletypesize{\scriptsize}
\setlength{\tabcolsep}{0.03in} 
\tablewidth{0pt}
\tablecaption{Predicted MIRAC Standard Star Fluxes on CWW system \label{stancal}}
\tablehead{
{(1)} & {(2)}  & {(3)}     & {(4)} & {(5)}  & {(6)}                       & {(7)}                       & {(8)}         & {(9)}  & {(10)}\\
{HD}  & {Alt.} & {Band}    & {Mag} & {unc.} & {F$_{\lambda}$}             & {unc.}                      & {F$_{\nu}$} & {unc.} & {unc.}\\
{Name}& {Name} & {\ldots}  & {\ldots} & {\ldots} & {(W\,cm$^{-2}$\,$\micron^{-1}$)}& {(W\,cm$^{-2}$\,$\micron^{-1}$)}& {(mJy)}  & {(mJy)}& {(\%)}
}
\startdata
1522  & $\iota$ Cet  & $L$      &  0.800 & 0.022 & 2.32E-15 & 4.99E-17 & 1.14E+05 & 2.46E+03 & 2.15\\
1522  & $\iota$ Cet  & $N$      &  0.807 & 0.021 & 4.81E-17 & 9.98E-19 & 1.77E+04 & 3.68E+02 & 2.08\\
1522  & $\iota$ Cet  & \eleven & 0.772 & 0.026 & 3.19E-17 & 9.01E-19 & 1.42E+04 & 4.02E+02 & 2.83\\
1522  & $\iota$ Cet  & \qs      &  0.775 & 0.030 & 6.02E-18 & 2.08E-19 & 6.23E+03 & 2.16E+02 & 3.46\\
12929 & $\alpha$ Ari & $L$      & -0.762 & 0.021 & 9.79E-15 & 2.01E-16 & 4.82E+05 & 9.88E+03 & 2.05\\
12929 & $\alpha$ Ari & $N$      & -0.754 & 0.020 & 2.02E-16 & 4.00E-18 & 7.46E+04 & 1.47E+03 & 1.98\\
12929 & $\alpha$ Ari & \eleven & -0.789 & 0.025 & 1.34E-16 & 3.70E-18 & 6.00E+04 & 1.65E+03 & 2.75\\
12929 & $\alpha$ Ari & \qs      & -0.787 & 0.030 & 2.54E-17 & 8.63E-19 & 2.62E+04 & 8.93E+02 & 3.40\\
29139 & $\alpha$ Tau & $L$      & -3.045 & 0.021 & 8.01E-14 & 1.62E-15 & 3.94E+06 & 7.99E+04 & 2.03\\
29139 & $\alpha$ Tau & $N$      & -3.013 & 0.020 & 1.62E-15 & 3.21E-17 & 5.97E+05 & 1.18E+04 & 1.98\\
29139 & $\alpha$ Tau & \eleven  & -3.074 & 0.025 & 1.10E-15 & 3.03E-17 & 4.92E+05 & 1.35E+04 & 2.75\\
29139 & $\alpha$ Tau & \qs      & -3.058 & 0.029 & 2.06E-16 & 6.91E-18 & 2.13E+05 & 7.14E+03 & 3.36\\
45348 & $\alpha$ Car & $L$      & -1.289 & 0.019 & 1.59E-14 & 3.04E-16 & 7.83E+05 & 1.50E+04 & 1.91\\
45348 & $\alpha$ Car & $N$      & -1.309 & 0.020 & 3.38E-16 & 6.53E-18 & 1.24E+05 & 2.41E+03 & 1.94\\
45348 & $\alpha$ Car & \eleven  & -1.307 & 0.025 & 2.17E-16 & 6.03E-18 & 9.67E+04 & 2.69E+03 & 2.78\\
45348 & $\alpha$ Car & \qs      & -1.307 & 0.029 & 4.10E-17 & 1.37E-18 & 4.24E+04 & 1.42E+03 & 3.35\\
48915 & $\alpha$ CMa & $L$      & -1.360 & 0.017 & 1.70E-14 & 2.96E-16 & 8.36E+05 & 1.46E+04 & 1.75\\
48915 & $\alpha$ CMa & $N$      & -1.348 & 0.018 & 3.50E-16 & 6.19E-18 & 1.29E+05 & 2.28E+03 & 1.77\\
48915 & $\alpha$ CMa & \eleven  & -1.346 & 0.023 & 2.24E-16 & 5.82E-18 & 1.00E+05 & 2.60E+03 & 2.59\\
48915 & $\alpha$ CMa & \qs      & -1.341 & 0.027 & 4.23E-17 & 1.36E-18 & 4.38E+04 & 1.41E+03 & 3.21\\
81797 & $\alpha$ Hya & $L$      & -1.362 & 0.019 & 1.70E-14 & 3.19E-16 & 8.37E+05 & 1.57E+04 & 1.88\\
81797 & $\alpha$ Hya & $N$      & -1.309 & 0.019 & 3.38E-16 & 6.39E-18 & 1.24E+05 & 2.36E+03 & 1.89\\
81797 & $\alpha$ Hya & \eleven  & -1.351 & 0.027 & 2.25E-16 & 6.59E-18 & 1.01E+05 & 2.94E+03 & 2.92\\
81797 & $\alpha$ Hya & \qs      & -1.350 & 0.030 & 4.26E-17 & 1.46E-18 & 4.41E+04 & 1.51E+03 & 3.42\\
106849& $\epsilon$ Mus & $L$    & -1.594 & 0.029 & 2.11E-14 & 5.83E-16 & 1.04E+06 & 2.87E+04 & 2.77\\
106849& $\epsilon$ Mus & $N$    & -1.647 & 0.027 & 4.61E-16 & 1.17E-17 & 1.70E+05 & 4.32E+03 & 2.54\\
106849& $\epsilon$ Mus & \eleven& -1.708 & 0.031 & 3.13E-16 & 1.01E-17 & 1.40E+05 & 4.53E+03 & 3.24\\
106849& $\epsilon$ Mus & \qs    & -1.700 & 0.039 & 5.89E-17 & 2.42E-18 & 6.09E+04 & 2.50E+03 & 4.10\\
108903& $\gamma$ Cru   & $L$    & -3.299 & 0.039 & 1.01E-13 & 3.71E-15 & 4.99E+06 & 1.83E+05 & 3.66\\
108903& $\gamma$ Cru   & $N$    & -3.354 & 0.038 & 2.22E-15 & 7.95E-17 & 8.18E+05 & 2.93E+04 & 3.58\\
108903& $\gamma$ Cru   & \eleven& -3.413 & 0.041 & 1.51E-15 & 6.13E-17 & 6.73E+05 & 2.74E+04 & 4.07\\
108903& $\gamma$ Cru   & \qs    & -3.403 & 0.043 & 2.83E-16 & 1.27E-17 & 2.92E+05 & 1.31E+04 & 4.49\\
128620& $\alpha$ Cen A & $L$    & -1.562 & 0.018 & 2.04E-14 & 3.58E-16 & 1.01E+06 & 1.76E+04 & 1.75\\
128620& $\alpha$ Cen A & $N$    & -1.564 & 0.018 & 4.27E-16 & 7.57E-18 & 1.57E+05 & 2.79E+03 & 1.77\\
128620& $\alpha$ Cen A & \eleven& -1.565 & 0.023 & 2.75E-16 & 7.13E-18 & 1.23E+05 & 3.18E+03 & 2.60\\
128620& $\alpha$ Cen A & \qs    & -1.566 & 0.027 & 5.20E-17 & 1.67E-18 & 5.38E+04 & 1.73E+03 & 3.21\\
133216& $\sigma$ Lib   & $L$    & -1.565 & 0.025 & 2.05E-14 & 5.00E-16 & 1.01E+06 & 2.46E+04 & 2.44\\
133216& $\sigma$ Lib   & $N$    & -1.619 & 0.022 & 4.49E-16 & 9.79E-18 & 1.66E+05 & 3.61E+03 & 2.18\\
133216& $\sigma$ Lib   & \eleven& -1.680 & 0.028 & 3.05E-16 & 9.03E-18 & 1.36E+05 & 4.03E+03 & 2.96\\
133216& $\sigma$ Lib   & \qs    & -1.670 & 0.036 & 5.73E-17 & 2.23E-18 & 5.92E+04 & 2.30E+03 & 3.89\\
135742& $\beta$ Lib    & $L$    &  2.874 & 0.019 & 3.44E-16 & 6.39E-18 & 1.69E+04 & 3.14E+02 & 1.86\\
135742& $\beta$ Lib    & $N$    &  2.899 & 0.019 & 7.00E-18 & 1.32E-19 & 2.58E+03 & 4.85E+01 & 1.88\\
135742& $\beta$ Lib    & \eleven&  2.904 & 0.025 & 4.48E-18 & 1.23E-19 & 2.00E+03 & 5.49E+01 & 2.75\\
135742& $\beta$ Lib    & \qs    &  2.915 & 0.029 & 8.40E-19 & 2.78E-20 & 8.68E+02 & 2.88E+01 & 3.32\\
150798& $\alpha$ TrA   & $L$    & -1.337 & 0.022 & 1.66E-14 & 3.50E-16 & 8.18E+05 & 1.72E+04 & 2.11\\
150798& $\alpha$ TrA   & $N$    & -1.329 & 0.021 & 3.44E-16 & 6.99E-18 & 1.27E+05 & 2.58E+03 & 2.03\\
150798& $\alpha$ TrA   & \eleven& -1.364 & 0.026 & 2.28E-16 & 6.38E-18 & 1.02E+05 & 2.85E+03 & 2.80\\
150798& $\alpha$ TrA   & \qs    & -1.361 & 0.030 & 4.31E-17 & 1.48E-18 & 4.46E+04 & 1.53E+03 & 3.44\\
167618& $\eta$ Sgr     & $L$    & -1.731 & 0.022 & 2.39E-14 & 5.08E-16 & 1.18E+06 & 2.50E+04 & 2.13\\
167618& $\eta$ Sgr     & $N$    & -1.696 & 0.021 & 4.82E-16 & 9.86E-18 & 1.78E+05 & 3.63E+03 & 2.04\\
167618& $\eta$ Sgr     & \eleven& -1.753 & 0.026 & 3.26E-16 & 9.13E-18 & 1.46E+05 & 4.08E+03 & 2.80\\
167618& $\eta$ Sgr     & \qs    & -1.786 & 0.031 & 6.37E-17 & 2.25E-18 & 6.59E+04 & 2.33E+03 & 3.53\\
216956& $\alpha$ PsA   & $L$    &  1.002 & 0.019 & 1.93E-15 & 3.62E-17 & 9.49E+04 & 1.78E+03 & 1.88\\
216956& $\alpha$ PsA   & $N$    &  1.001 & 0.019 & 4.02E-17 & 7.65E-19 & 1.48E+04 & 2.82E+02 & 1.90\\
216956& $\alpha$ PsA   & \eleven&  1.004 & 0.025 & 2.58E-17 & 7.11E-19 & 1.15E+04 & 3.18E+02 & 2.76\\
216956& $\alpha$ PsA   & \qs    &  1.008 & 0.029 & 4.86E-18 & 1.62E-19 & 5.02E+03 & 1.67E+02 & 3.33\\
\enddata
\tablecomments{All stated quantities assume that a KRS-5 dewar window is used. If the KBr window is
used, the values are nearly identical (to within 7\% of the stated uncertainties).}
\end{deluxetable}

\end{document}